\documentclass[trans,comsoc]{IEEEtran}
\usepackage{cite}
\usepackage{amsmath,amssymb,amsfonts}
\usepackage{algorithmic}
\usepackage{graphicx}
\usepackage{textcomp}
\usepackage{epstopdf}
\usepackage{stfloats}
\usepackage{setspace}
\ifCLASSOPTIONcompsoc
\usepackage[caption=false, font=normalsize, labelfont=sf, textfont=sf]{subfig}
\else
\usepackage[caption=false, font=footnotesize]{subfig}
\usepackage{bm}

\usepackage[T1]{fontenc}
\usepackage{cite}
\usepackage{amsmath,amssymb,amsfonts}
\usepackage{algorithmic}
\usepackage{graphicx}
\usepackage{textcomp}
\usepackage{xcolor}
\usepackage{bm}
\usepackage{changepage}
\usepackage{amsmath}
\usepackage{amstext}
\usepackage{stfloats}

\usepackage{amsmath}
\def\BibTeX{{\rm B\kern-.05em{\sc i\kern-.025em b}\kern-.08em
    T\kern-.1667em\lower.7ex\hbox{E}\kern-.125emX}}

\usepackage{hyperref}
\hypersetup{
hidelinks,
colorlinks=true,
linkcolor=black,
citecolor=black,
urlcolor = black
}

\begin{document}

\title{Wideband Beamforming for RIS Assisted Near-Field Communications
			\author{Ji Wang, \IEEEmembership{Senior Member,~IEEE}, Jian Xiao, Yixuan Zou, \IEEEmembership{Member,~IEEE}, Wenwu Xie, \\
			and Yuanwei Liu, \IEEEmembership{Fellow,~IEEE}}
			    \thanks{This work was supported in part by the National Natural Science Foundation of China under Grant 62101205, and in part by the Key Research and Development Program of Hubei Province under Grant 2023BAB061, in part by the Fundamental Research Funds for the Central Universities of China under grant CCNU24ai004, and in part by the Graduate Innovation Program of Central China Normal University under Grant 2024CXZZ151. \emph{(Corresponding author: Wenwu Xie.)}}
    \thanks{Ji Wang and Jian Xiao are with the Department of Electronics and Information Engineering, College of Physical Science and Technology, Central China Normal University, Wuhan 430079, China (e-mail: jiwang@ccnu.edu.cn; jianx@mails.ccnu.edu.cn).}
	\thanks{Yixuan Zou is with the School of Electronic Engineering and Computer Science, Queen Mary University of London, E1 4NS London, U.K. (e-mail: yixuan.zou@qmul.ac.uk).}
		\thanks{Wenwu Xie is with the School of Information Science and Engineering, Hunan Institute of Science and Technology, Yueyang 414006, China (e-mail: gavinxie@hnist.edu.cn).}
				\thanks{Yuanwei Liu is with the Department of Electrical and Electronic Engineering, The University of Hong Kong, Hong Kong (e-mail: yuanwei@hku.hk).}
		}
\maketitle
\begin{abstract}
A near-field wideband beamforming scheme is investigated for reconfigurable intelligent surface (RIS) assisted multiple-input multiple-output (MIMO) systems, in which a deep learning-based end-to-end (E2E) optimization framework is proposed to maximize the system spectral efficiency. To deal with the near-field double beam split effect, the base station is equipped with frequency-dependent hybrid precoding architecture by introducing sub-connected true time delay (TTD) units, while two specific RIS architectures, namely true time delay-based RIS (TTD-RIS) and virtual subarray-based RIS (SA-RIS), are exploited to realize the frequency-dependent passive beamforming at the RIS. Furthermore, the efficient E2E beamforming models without explicit channel state information are proposed, which jointly exploits the uplink channel training module and the downlink wideband beamforming module. In the proposed network architecture of the E2E models, the classical communication signal processing methods, i.e., polarized filtering and sparsity transform, are leveraged to develop a signal-guided beamforming network. Numerical results show that the proposed E2E models have superior beamforming performance and robustness to conventional beamforming benchmarks. Furthermore, the tradeoff between the beamforming gain and the hardware complexity is investigated for different frequency-dependent RIS architectures, in which the TTD-RIS can achieve better spectral efficiency than the SA-RIS while requiring additional energy consumption and hardware cost.
\end{abstract}
\begin{IEEEkeywords}
	Deep learning, near-field communications, reconfigurable intelligent surface, wideband beamforming.
\end{IEEEkeywords}

%
\IEEEpeerreviewmaketitle

\section{Introduction}
\IEEEPARstart{T}{he} sixth-generation (6G) wireless networks aim to further deliver high throughput, achieve massive connectivity, and enhance energy efficiency. In order to accomplish these promising objectives, extremely large scale antenna arrays (ELAAs) and tremendously high frequencies form a pair of prospective technological solutions. In particular, as a new type of metamaterial antenna, reconfigurable intelligent surface (RIS) technology has been regarded as one of the highly anticipated candidate ELAA solution to construct a smart radio environment \cite{SRE, 9410457}. In this case, the near-field boundary in 6G communications will be significantly extended due to the increase of Rayleigh distance that is positively correlated with the array aperture and the communication frequency \cite{cui2022near}. Considering a large number of available bandwidth in high frequencies, e.g., millimeter wave (mmWave) and Terahertz (THz), the near-field wideband RIS communications is becoming {an} up-and-coming communication paradigm in 6G era \cite{10380596}.

In near-field wideband RIS systems, new electromagnetic (EM) characteristics need to be considered compared to the classic far-field narrowband systems. Firstly, in contrast to the planar wavefront assumption in the far-field channel modeling, the near-field channel involves both angle and distance dimensions due to the spherical wavefront in the near-field radiation, which results in the near-field \emph{beamfocusing} effect instead of the far-field beamsteering \cite{Liu2023Near, 10135096}. Secondly, the near-field wideband channels can be strongly \emph{frequency-dependent} due to the large bandwidth between different subcarriers. However, for the popular hybrid beamforming architecture in ELAA systems, the typical analog beamformer is \emph{frequency-independent}. Consequently, {the beams generated in different frequencies may be focused at different locations,} which is termed as the \emph{beam split} effect \cite{Dai2022Delay}. Especially, in RIS enabled wideband communications, since reflection units at the RIS only carry out the passive phase shifting operation, the impinging and reflected beams at the RIS will be also split into different physical directions for different frequencies. In this case, the specific property of the frequency-independent analog precoding at the base station (BS) and phase shifting at the RIS cause the unique \emph{double beam split} effect \cite{Su2023Wideband}. Consequently, efficient frequency-dependent beamforming architectures and optimization schemes are urgently expected to investigate for near-field wideband RIS systems.


\subsection{Prior works}
\subsubsection{{Wideband RIS Communications}} To deal with the beam split effect in RIS-aided mmWave/THz systems, distributed RISs and delay adjustable RISs are two feasible solutions. In \cite{yan2023}, the distributed RIS deployment strategy was proposed to relieve the beam split effect, which required the high deployment cost and still relied on the frequency-independent phase shifting architecture \cite{Hao2022Ultra}. In order to completely break through this limitation of the analog phase shifting circuit at the RIS, the proposed true time delay (TTD) module in the wideband hybrid precoder architecture was extended to the classic RIS architecture. Specifically, in \cite{Sun2022Time}, the RIS element can realize the frequency-selective operation by introducing TTD units, in which a sub-surface architecture of RIS was provided to balance the power gain and the hardware cost. Furthermore, the authors of \cite{Su2023Wideband} proposed a sub-connected phase-delay-phase RIS architecture to provide an energy-efficient implementation for the frequency-dependent wideband phase shifting scheme. Moreover, the emerging simultaneously transmitting and reflecting RIS (STAR-RIS) architecture has been explored in THz wideband communications \cite{9570143, Wang2023STAR}.
{\subsubsection{{Near-Field Wideband ELAA Communications}} In near-field wideband communications, the intricacies of near-field beamfocusing and wideband beam split effects are deeply coupled, significantly complicating beamforming optimization efforts. The authors of \cite{10541333} investigated the near-field beam split phenomenon for conventional ELAA systems, in which a phase-delay focusing approach was proposed to improve effective beamforming gain. Furthermore, in \cite{10458958}, two distinct TTD configurations, i.e., a serial TTD and a hybrid serial-parallel TTD, were proposed to address the spatial-wideband effect in near-field ELAA systems. In \cite{10505154}, focusing on holographic metasurface antennas-based ELAA systems, a multi-user beam combining optimization framework was proposed, which accounted for both the near-field and dual-wideband effects in holographic communication systems.}
\subsubsection{{Near-Field Wideband RIS Communications}}
For RIS enabled near-field wideband systems, {
the authors of \cite{10283765} proposed an RIS configuration approach to maximize the communication rate, which effectively mitigates the beamforming losses caused by the near-field beam split effect.} In \cite{Hao2023The}, the delay adjustable metasurface technique in \cite{an2021reconfigurable} that can adjust the delays of signals reflected by different {RIS} elements was applied to alleviate the beam split effect. However, the fully-connected TTD module in \cite{Hao2023The} will lead to excessive hardware cost and power consumption, which requires the number of delay units has to be equal to the number of massive RIS elements. Considering the uplink achievable rate optimization in THz RIS systems, the authors of \cite{Cheng2023Achievable} divided the RIS into multiple virtual subarrays, in which the phase shift of each subarray was optimized according to the corresponding subcarrier channel. In this way, the RIS was endowed with the ability to carry out the frequency-dependent passive beamforming at different sub-bandwidths. However, the effective aperture of RIS will be shrunk for the virtual subarray architecture, which results in the significant energy loss of received signals at the RIS and hence restricts the system performance.

\subsection{Motivations and Contributions}  
While a few of research efforts have been devoted to investigate the near-field wideband RIS systems, {the necessary prior assumptions, such as the known array manifold \cite{Hao2023The} and the line-of-sight (LOS)-dominant channel \cite{Cheng2023Achievable}, were required for the RIS phase shifting derivation. In addition, in the aforementioned works, the authors only focus on the phase shifting design at the RIS, while the BS was assumed to be equipped with the single antenna or the predetermined fully-digital precoder. Consequently, the comprehensive solution is expected to be further investigated for the joint passive and active beamforming optimization}, which involves the coupled non-convex optimization in the wideband hybrid beamforming at the BS and phase shifting at the RIS. Moreover, when the RIS and the BS are equipped with large-scale antenna arrays, the required high-dimensional channel acquisition is also an intractable challenge for the beamforming optimization.

Against the above background, in this work, we investigate the near-field wideband beamforming design for RIS-aided multiple-input multiple-output orthogonal frequency division multiplexing (MIMO-OFDM) systems. Our main contributions are summarized as follows.

\hangafter=1
\setlength{\hangindent}{2em}
$\bullet$ {We investigate frequency-dependent hybrid precoding and phase shifting architecture for near-field wideband RIS systems, aiming for alleviating the beamforming performance loss caused by the near-field double beam split effect.} Specifically, the BS is equipped with frequency-dependent hybrid precoding architecture by introducing sub-connected TTD units, to deal with the near-field beam split effect at the BS. Furthermore, considering the wideband beam split effect at the RIS, two specific RIS architectures, namely true time delay-based RIS (TTD-RIS) and virtual subarray-based RIS (SA-RIS), are exploited to realize the frequency-dependent passive beamforming at the RIS.

\setlength{\hangindent}{2em}
$\bullet$ We propose a deep learning-based end-to-end (E2E) beamforming optimization framework {to maximize the effective spectral efficiency in RIS-aided MIMO-OFDM systems.} {The proposed E2E model is composed of the uplink channel training (UL-CT) module and the downlink beamforming (DL-BF) module,} in which the learnable combining matrix at the BS and phase shifting at the RIS are designed to realize the joint optimization of beamforming and channel estimation {with limited pilot overhead}. In contrast to the pre-defined combining matrix and reflection pattern in traditional channel estimators, the combining matrix and phase shifting in the proposed UL-CT module can be adaptively tuned according to dynamic wireless environments. 

\setlength{\hangindent}{2em}
$\bullet$ We exploit an efficient signal-guided beamforming network architecture based on the proposed E2E optimization framework, which integrates advanced neural network architectures and classical communication signal processing methods. Specifically, in the proposed UL-CT module, we design a polar attention architecture to imitate the typical communication signal filtering {in} the frequency domain and time-spatial domain, which {can finely learn effective} latent channel semantic information from the received pilots. Motivated by the natural channel sparsity for high-frequency ELAA systems, a learnable discrete Fourier transform (DFT) is introduced into the proposed DL-BF module, which guides and accelerates the convergence of the beamforming network. 

\setlength{\hangindent}{2em}
$\bullet$ Our numerical results reveal that a superior beamforming performance can be achieved by the proposed E2E models over the conventional beamforming benchmarks. Specifically, compared to the conventional hybrid precoding and the classic RIS architecture, the proposed TTD-RIS and SA-RIS can effectively mitigate the near-field double beam split effect. Furthermore, the proposed E2E models can jointly optimize the active and passive beamforming with the implicit CSI, which reduces the required training overhead and improves the effective spectral efficiency. Moreover, the robustness and generalization of the proposed E2E models are evaluated under various system setups.

\subsection{Organizations and Notations}  

The remainder of this paper is organized as follows. Section II introduces the near-field wideband channel modeling and system model in RIS assisted MIMO-OFDM systems. In Section III,  the deep learning-based near-field wideband beamforming framework is proposed. Furthermore, the signal-guided network architecture is presented in Section IV. Section V provides numerical results of the proposed E2E models. In Section VI, this paper is comprehensively summarized.

\emph{Notations}: Lower-case and upper-case boldface letters denote a vector and a matrix, respectively; $\mathbf{A}^T$ and $\mathbf{A}^H$ denote the transpose and conjugate transpose of matrix $A$, respectively; $a^*$ denotes the conjugate of complex number $a$; $\rm{diag}(\mathbf{a})$ denotes the diagonal matrix with the vector $\mathbf{a}$ on its diagonal; ${\bf{I}}_a$ is a $a \times a$ identity matrix, while ${\bf{1}}_a$ is a $a \times 1$ vector, satisfying ${\bf{1}}_i=1, \forall i=\{1, \ldots, a\}$; Symbols $| {\cdot}|$, $\left\| {\cdot} \right\|$, and $\left\| {\cdot} \right\|_{F}$ denote the $\ell_1$, $\ell_2$, and Frobenius norm, respectively; $\propto$ denotes the proportionality relation. $\odot$ and $ \otimes $ denote the Hadamard product and Kronecker product, respectively. $\Re(\mathbf{A})$ and $\Im(\mathbf{A})$ denote the real and imaginary components of the complex-value matrix $\mathbf{A}$. $\mathrm{det}(\mathbf{A})$ denote the determinant of the matrix $\mathbf{A}$. Symbol $\mathrm j$ denotes the imaginary number, and $\left\lfloor{\cdot} \right\rfloor$ denotes the round down operation.

\section{System Model and Problem Formulation}
\begin{figure*}[t]
	\centerline{\includegraphics[width=6.0in]{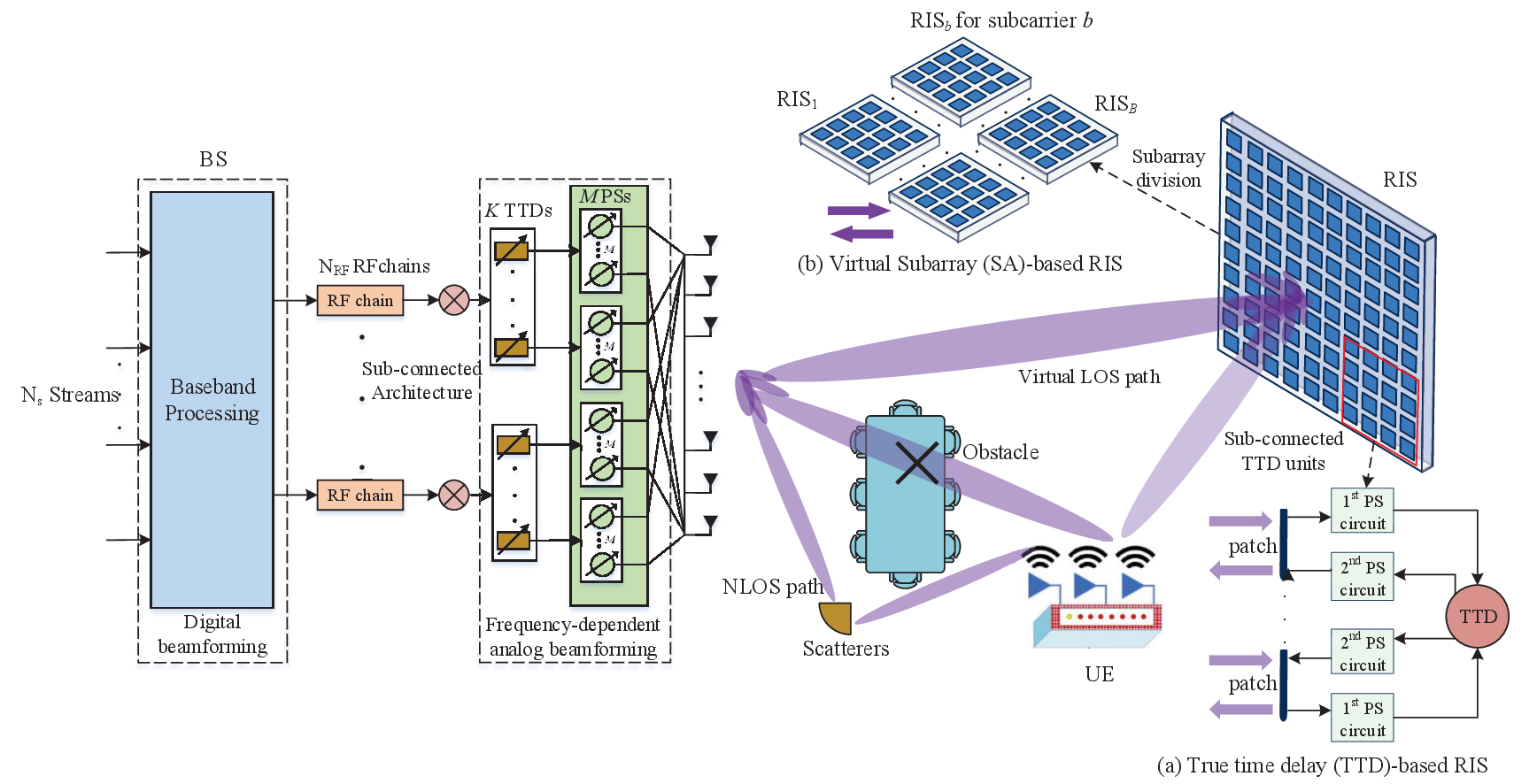}}
	\caption{{Near-field wideband systems assisted by different frequency-dependent RIS architectures. (a) The TTD-RIS architecture; (b) The SA-RIS architecture.}}
	\label{fig1}
\end{figure*}
As shown in Fig.~\ref{fig1}, we consider an $N$-element RIS assisted mmWave MIMO-OFDM systems with $M$ transmit and $U$ receive antennas. Both transmit antennas at the BS and the receive antennas at the user equipment (UE) are placed in uniform linear arrays (ULAs), while the reflection elements at the RIS are placed in uniform plane arrays (UPAs), i.e., $N=N_1 \times N_2$. Considering the high susceptibility in mmWave communications and practical complex wireless environments, we assume that the LOS path between the BS and the UE is completely blocked, while the non-LOS (NLOS) paths are constructed by limited clustered scatterers. To enhance mmWave  communications, the RIS is introduced to provide the virtual LOS path between the BS and the UE. 


To deal with the double beam split effect caused by the coupling of wideband beam split at the BS and the RIS, we introduce the newly fashionable frequency-dependent beamforming architecture. Firstly, the sub-connected TTD module is introduced to the hybrid beamforming architecture at the BS, which is composed of $K$ TTDs and $M_\text{RF} \ll M$ radio frequency (RF) chains. Specifically, each RF chain is connected to $K$ TTDs, and then each TTD is connected to $P=M/K$ phase-shifters (PSs). By utilizing the TTD module, the traditional analogy beamforming will evolve into the frequency-dependent analogy beamforming at the BS. Then, for the wideband RIS architecture, we provide two feasible candidate solutions, i.e., sub-connected TTD-RIS in Fig.~\ref{fig1}(a) and virtual SA-RIS architecture in Fig.~\ref{fig1}(b), whose detailed design guidelines are presented in Section II.C.

\subsection{Near-Field Boundary in RIS-Aided MIMO Systems}
In near-field communications, the Rayleigh distance $R$ is a widely used criterion to define the near-field boundary, which is given by \cite{kraus2002antennas}
\begin{align}\label{MISO}
R = \frac{2D^2}{\lambda},
\end{align}
where $D$ and $\lambda$ denote the antenna aperture and the wavelength, respectively. In multiple-input single-output (MISO) systems, the near-field boundary can be defined according to \eqref{MISO}. With the increase of antenna array and communication frequency, the near-field region will be significantly extended.

In particular, for RIS-aided MIMO communications, the near-field criterion is different for the BS$\to$UE direct link and the BS$\to$RIS$\to$UE cascaded link due to different EM characteristics. For the BS$\to$UE MIMO communications, the near-field boundary is given by \cite{cui2022near}
\begin{align}\label{MIMO}
R = \frac{2(D^\text{B}+D^\text{U})^2}{\lambda},
\end{align}
where $D^\text{B}$ and $D^\text{U}$ denote the antenna aperture of the BS and the UE, respectively. 

In contrast to MISO/MIMO channels in conventional communication systems, the cascaded BS$\to$RIS$\to$UE reflection link comprises BS$\to$RIS and RIS$\to$UE links in RIS systems. According to the near-field criterion in \cite{cui2022near}, the near-field region in RIS systems can be expressed as
\begin{equation}\label{RIS_NF}
\begin{split}
\frac{{{r_1}{r_2}}}{{r_1 + r_2}} <  \frac{{2{D^2}}}{\lambda },
\end{split}
\end{equation}
where $r_1$ and $r_2$ denote the BS$\to$RIS and RIS$\to$UE distance, respectively. We observe that the near-field region in RIS systems is determined by the harmonic mean of $r_1$ and $r_2$. In other words, as long as any of $r_1$ and $r_2$ is shorter than Rayleigh distance, the near-field communications will occur in RIS systems due the passive property of the RIS.

\subsection{Near-Field Wideband Channel Model}
\begin{figure}[t]
	\centerline{\includegraphics[width=3.0in]{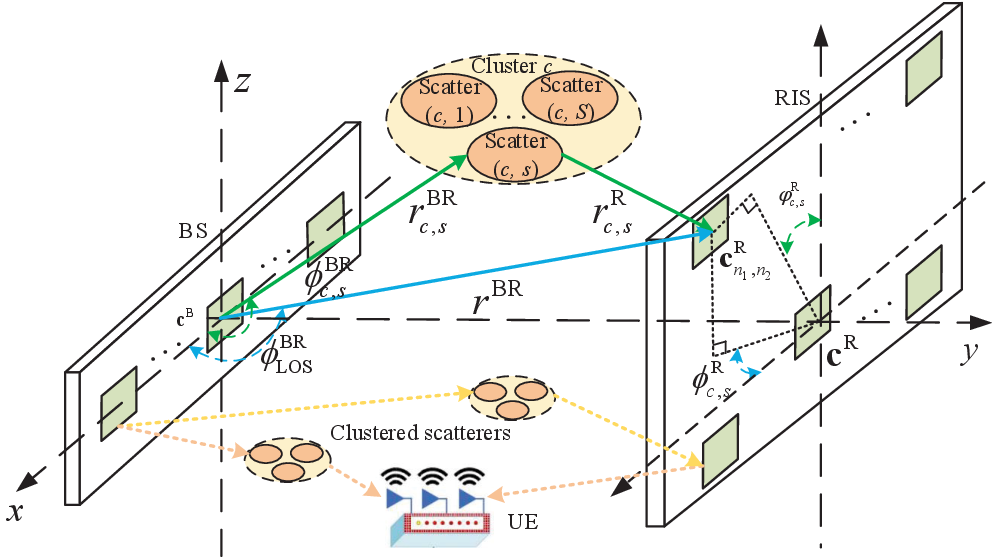}}
	\caption{System layout for the near-filed channel modeling in BS$\to$RIS link.}
	\label{NF_Channel}
\end{figure}

{Considering the three-dimensional Cartesian coordinate system in Fig.~\ref{NF_Channel}, we assume that both BS and RIS lie on the $xz$-plane,} whose array center coordinate are set to $\mathbf{c}^\text{B}=\left( {x^\text{B},{y^\text{B}},{z^\text{B}}} \right)$ and $\mathbf{c}^\text{R}=\left( {{x^\text{R}},y^\text{R},{z^\text{R}}} \right)$, respectively. The coordinate of $\text{UE}$ $\mathbf{c}^\text{U}=\left( {x^\text{U},y^\text{U},z^\text{U}} \right)$ is randomly distributed around the RIS. Let $\Delta m$, $\Delta {u}$ and $\Delta n = \Delta {n_1} = \Delta {n_2}$ denote the distance between two adjacent antennas (elements) at the BS, the UE, and the RIS, respectively. Generally, the antenna spacing is set to $d = \Delta m= \Delta u = \Delta n=\frac{c}{2f_c}$ in large-scale array communications, where $c$ and $f_c$ denote the speed of light and the central carrier frequency, respectively. Hence, the coordinate of the AP antenna ${{m}}$ is $\mathbf{c}^\text{B}_{m}=\left( x^\text{B} + ({m} - \frac{{{M} + 1}}{2})d, {y^\text{B}}, {z^\text{B}} \right)$. Accordingly, the coordinate of the UE antenna ${{u}}$ is $\mathbf{c}^\text{U}_{u}=\left( x^\text{U} + ({u} - \frac{{{U} + 1}}{2})d, {y^\text{U}}, {z^\text{U}} \right)$. For the UPA-based RIS, the coordinate of the RIS element $\left( {{n_1},{n_2}} \right)$ is $\mathbf{c}^\text{R}_{n_1,n_2}=\left({x^\text{R}} + ({n_1} - \frac{{{N_1} + 1}}{2})d, {{y^\text{R}}, {z^\text{R}} + ({n_2} - \frac{{{N_2} + 1}}{2})d} \right)$. In this work, we consider clustered scatterer propagation environments \cite{basar2021indoor}, in which the coordinate of scatterer $s$ $\left( {1 \le s \le {S_c}} \right)$ in cluster $c$ $\left( {1 \le c \le C_\text{s}} \right)$ is denoted as $\mathbf{c}^\text{S}_{c,s}=\left( {{x^\text{S}_{c,s}},{y^\text{S}_{c,s}},{z^\text{S}_{c,s}}} \right)$.

According to the wideband ray-tracing-based channel model \cite{Tarboush2021TeraMIMO}, the BS$\to$UE NLOS channel $\mathbf{D}[\tau] \in \mathbb {C}^{U \times M} $ at the $\tau$-th delay can be expressed as
\begin{align}
\mathbf{D}[\tau] ={{\gamma^\text{BU} }\sum\limits_{c=1}^{{{C_\text{s}}}}{\sum\limits_{s=1}^{{{{{S}_{c}}}}}{{{{\varsigma }_{c,s}^{\text{BU}}}}}}\sqrt{G_\text{B}G_\text{U}L_{c,s}^\text{BU}}\mathbf{u}^\text{BU}_{c,s}{(\mathbf{a}_{c,s}^{\text{BU}})^T}}\delta (\tau T_s  - {\tau _{c,s}^{\text{BU}}}),
\end{align}
where ${\gamma^\text{BU}}=\sqrt{\frac{1}{\sum\nolimits_{c=1}^{{C_\text{s}}}{{{{{S}}}_{c}}}}}$ is a normalization factor, ${{{\varsigma }}_{c, s}^{\text{BU}}}\sim \mathcal{C}\mathcal{N}(0, 1)$ is the complex gain, and $L_{c, s}^\text{BU}$  is the path loss for scatterer $(c, s)$. Parameters $G_\text{B}$ and $G_\text{U}$ denote the antenna gain at the transmitter ans receiver, respectively. The function $\delta(\tau)$ denotes the dirac function for $T_s$-spaced signaling evaluated at $\tau$ seconds. $\mathbf{a}_{c,s}^{\text{BU}} \in {\mathbb{C}^{M \times 1}}$ denotes the transmitting array response at the BS, and $\mathbf{u}_{c,s}^\text{SU} \in {\mathbb{C}^{U \times 1}}$ represents the receiving response at the UE. Parameter $\tau _{c,s}^{\text{BU}}$ denotes the path delay of scatterer $(c,s)$.

Suppose system bandwidth $W$ is divided into $B$ orthogonal subcarriers in OFDM systems, the communication frequency at subcarrier $b$ can be expressed as $f_b=f_c+\frac{W(2b-1-B)}{2B},(1\le b \le B)$. In this work, we adopt the uniform spherical wave model to characterize the near-field wideband array response vector \cite{Liu2023Near}. The ULA array response at subcarrier frequency $f_b$ for BS$\to$scatterer $(c,s)$ is given by
 \begin{flalign}\label{nearULA}
&\mathbf{a}(f_b,\phi _{c,s}^\text{BU}, r^\text{BU}_{c,s})=e^{-j \frac{2 \pi f_b}{c}\left(-m d \cos \phi _{c,s}^\text{BU} +\frac{{{m}}^{2} d^{2}(\sin \phi _{c,s}^\text{BU} )^2}{2 r^\text{BU}_{c,s}}\right)},
\end{flalign}
where $\phi _{c,s}^\text{BU}$ denotes the azimuth angle of departure (AoD) for scatterer $(c, s)$ at the BS, and $r^\text{BU}_{c,s}=\left\| {\mathbf{c}^\text{B}-\mathbf{c}^\text{S}_{c,s}} \right\|$ is the distance between the BS and scatterer $(c,s)$ in the BS$\to$UE link. 

Since the number of antennas of the UE is generally limited, the far-field radiation with the planner wavefront can be directly adopted for the channel modeling between scatterer $(c,s)$ and the UE in the BS$\to$UE link. In this case, the ULA array response $\mathbf{u}$ for scatterer $(c,s)$ is given by
 \begin{flalign}
{\mathbf{u}(f_b,\vartheta _{c,s}^\text{BU})=\left[1,  e^{{-j \frac{2 \pi f_b}{c}}d\sin \vartheta _{c,s}^\text{BU}}, \ldots, e^{{-j \frac{2 \pi f_b}{c}}(U-1)d\sin \vartheta _{c,s}^\text{BU}}\right]^T},
\end{flalign}
where $\vartheta _{c,s}^\text{BU}$ denotes the angle of arrival (AoA) for scatterer $(c, s)$ at the UE. 


In OFDM systems, the time-domain signal is transformed into the frequency domain by utilizing the DFT. Correspondingly, the frequency-domain BS$\to$UE NLOS channel $\mathbf{D}[b] \in \mathbb {C}^{U \times M} $ at subcarrier $b$ is given by \cite{Tarboush2021TeraMIMO}
\begin{align}
\mathbf{D}[b] ={{\gamma^\text{BU} }\sum\limits_{c=1}^{{{C_\text{s}}}}{\sum\limits_{s=1}^{{{{{S}_{c}}}}}{{{{\varsigma }_{c,s}^\text{BU}}}}}\sqrt{G_\text{B}G_\text{U}L_{c,s}^{{\text{BU}}}}\mathbf{u}^\text{BU}_{c,s}{(\mathbf{a}_{c,s}^\text{BU})^{T}}}e^{\frac{-j2\pi bW \tau _{c,s}^\text{BU}}{B}}.
\end{align}

The BS$\to$RIS channel $\mathbf{G}[\tau]=\mathbf{G}_{\text{LOS}}[\tau]+\mathbf{G}_{\text{NLOS}}[\tau]\in \mathbb {C}^{N \times M}$ is composed of a stable LOS path and clustered NLOS paths. In Fig.~\ref{NF_Channel}, the specific near-field BS$\to$RIS channel modeling is presented, wherein the central antenna of the BS is set to the origin of the coordinate system. The LOS channel $\mathbf{G}_{\text{LOS}}[\tau]$ between the BS and the RIS at the $\tau$-th delay can be expressed as
\begin{align}\label{UPARIS}
 \mathbf{G}_{\text{LOS}}[\tau] = &\bar {\varsigma}^{\text{BR}}_{\text{LOS}}\mathbf{b}_{\text{LOS}}^\text{BR}{(\mathbf{a}}_{\text{LOS}}^\text{BR})^{T} \odot \left(\mathbf{G}^x_c \otimes \mathbf{G}^z_c\right)\delta (\tau T_s  - {\tau_\text{LOS}^\text{BR}}),
\end{align}
where parameter $\bar {\varsigma}^{\text{BR}}_{\text{LOS}} = \sqrt{G_\text{B}G_\text{R}L_{\text{LOS}}^\text{BR}}$ is composed of transmitting antenna gain, RIS element gain, and path loss. ${\tau_\text{LOS}^\text{BR}}$ denotes the path delay of the LOS channel. $\mathbf{a}_{\text{LOS}}^\text{BR} \in {\mathbb{C}^{M \times 1}}$ and $\mathbf{b}_{\text{LOS}}^\text{BR} \in {\mathbb{C}^{N \times 1}}$ denote the transmitting array response at the BS and the receiving response at the RIS for the BS$\to$RIS LOS path, respectively. Due to the space limitation, the specific definition of $\mathbf{a}_{\text{LOS}}^\text{BR}$ at the BS can refer to \eqref{nearULA}. The near-field UPA array response $\mathbf{b}_{\text{LOS}}^\text{BR}$ at the RIS is given by
\begin{subequations}\label{nearUPA}
\begin{flalign}
&\mathbf{b}_{\text{LOS}}^\text{BR} =\mathbf{b}_{x}(f_b,\phi^\text{R}_\text{LOS}, \varphi^\text{R}_\text{LOS}, r^\text{BR}) \otimes \mathbf{b}_{z}(f_b,\varphi^\text{R}_\text{LOS}, r^\text{BR}), \label{Zb} \\
&{\left[\mathbf{b}_{x}\right]_{{n}_1}=e^{-j \frac{2 \pi f_b}{c}\left(-\bar{n}_1 d \zeta^\text{R}_\text{LOS} +\frac{{\bar{n}_1}^{2} d^{2}\left(1-(\zeta^\text{R}_\text{LOS} )^2 \right)}{2 r^\text{BR}_{c,s}}\right)}}, \label{Zc} \\
&{\left[\mathbf{b}_{z}\right]_{{n}_2}=e^{-j \frac{2 \pi f_b}{c}\left(-\bar{n}_2 d \cos \varphi _{c,s}^\text{R}+\frac{{\bar{n}_2}^{2} d^{2} \sin ^{2} \varphi _{c,s}^\text{R}}{2 r^\text{BR}_{c,s}}\right)}, \label{Zd}} 
\end{flalign}
\end{subequations}
where $\phi^\text{R}_\text{LOS}$ and $\varphi^\text{R}_\text{LOS}$ denote the azimuth and elevation angle of AoA of LOS path at the RIS, respectively. Parameters $\bar{n}_1 = {n_1} - \frac{{{N_1} + 1}}{2}$, $\bar{n}_2= {n_2} - \frac{{{N_2} + 1}}{2}$, $\zeta^\text{R}_\text{LOS} =\cos \phi^\text{R}_\text{LOS} \sin \varphi^\text{R}_\text{LOS} $, and $r^\text{BR}=\left\| {\mathbf{c}^\text{B}-\mathbf{c}^\text{R}} \right\|$ denotes the distance between the BS and the RIS. Note that the near-field LOS BS$\to$RIS channel in \eqref{UPARIS} includes an additional coupled component, i.e., $\mathbf{G}^x_c \otimes \mathbf{G}^z_c$, which can be expressed as
\begin{subequations}\label{couple}
\begin{flalign}
&\left[\mathbf{G}^x_c \right]_{{n}_2} = e^{-j \frac{2 \pi f_b}{c r^\text{BR}}\bar{n}_2 d^2\left(1-\zeta^2_\text{LOS}\right)}, \label{Zf} \\
&\left[\mathbf{G}^z_c \right]_{{n}_1,m} = e^{-j \frac{2 \pi f_b}{c r^\text{BR}}\bar{n}_1 m d^2\sin^2 \varphi^\text{BR}_\text{LOS}}. \label{Zg}
\end{flalign}
\end{subequations}
Due to the presence of the above coupled component, near-field LOS MIMO channels exhibit higher degree of freedoms (DoFs) than far-field LOS MIMO channels. Hence, the higher near-field DoFs can be exploited by constructing the virtual LOS path in RIS systems.


For the clustered NLOS components of the BS$\to$RIS channel, the NLOS channel $\mathbf{G}_{\text{NLOS}}[\tau]$ at the $\tau$-th delay can be expressed as
\begin{align}
\mathbf{G}_{\text{NLOS}}[\tau] ={{\gamma^\text{BR} }\sum\limits_{c=1}^{{{C_\text{s}}}}{\sum\limits_{s=1}^{{{{{S}_{c}}}}}}\bar {\varsigma}^{\text{BR}}_{c,s}\mathbf{b}^{\text{BR}} _{c,s}({\mathbf{a}_{c,s}^\text{BR}})^{T}}\delta (\tau T_s  - {\tau^{\text{BR}} _{c,s}}),
\end{align}
where ${\gamma^\text{BR}}$ is a normalization factor for scatterer paths, and parameter $\bar {\varsigma}^{\text{BR}}_{c,s} = {\varsigma }^{\text{BR}}_{c,s}\sqrt{G_\text{B}G_\text{R}L_{c,s}^\text{BR}}$ is composed of the complex gain, the array gain and the path loss for scatterer $(c, s)$ in the BS$\to$RIS link. $\mathbf{a}^\text{BR}_{c,s} \in {\mathbb{C}^{M \times 1}}$ denotes the transmitting array response at the BS for the BS$\to$RIS NLOS path, and $\mathbf{b}^\text{BR}_{c,s} =\mathbf{b}_{x}(f_b,\phi^\text{R}_{c,s}, \varphi^\text{R}_\text{c,s}, r^\text{R}_{c,s}) \otimes \mathbf{b}_{z}(f_b,\varphi^\text{R}_{c,s}, r^\text{R}_{c,s})\in {\mathbb{C}^{N \times 1}}$ denotes the receiving response at the RIS. Parameters $\phi^\text{R}_{c,s}$ and $\varphi^\text{R}_{c,s}$ denote the azimuth and elevation angle of AoA for scatterer $(c,s)$ at the RIS, respectively. $r^\text{R}_{c,s}=\left\| {\mathbf{c}^\text{R}-\mathbf{c}^\text{S}_{c,s}} \right\|$ and $\tau^{\text{BR}} _{c,s}$ denotes the path delay of scatterer $(c,s)$ in the BS$\to$RIS link. 

By carrying out the DFT on $\mathbf{G}[\tau]$, the frequency-domain channel $\mathbf{G}[b]=\mathbf{G}_{\text{LOS}}[b]+\mathbf{G}_{\text{NLOS}}[b]$ at the $b$-th subcarrier is given by
\begin{align}
\mathbf{G}[b]&{=}\underbrace{\bar {\varsigma}^{\text{BR}}_{\text{LOS}}\mathbf{b}_{\text{LOS}}^\text{BR}({\mathbf{a}}_{\text{LOS}}^\text{BR})^{T} \odot \left(\mathbf{G}^x_c \otimes \mathbf{G}^z_c\right)e^{\frac{-j2\pi bW {\tau_\text{LOS}^\text{BR}}}{B}}}_{\mathbf{G}_{\text{LOS}}[b]} \nonumber \\
&{+}\underbrace{{{\gamma^\text{BR} }\sum\limits_{c=1}^{{{C_\text{s}}}}{\sum\limits_{s=1}^{{{{{S}_{c}}}}}{{{{\varsigma }^{\text{BR}} _{c,s}}}}}\bar {\varsigma}^{\text{BR}}_{c,s}\mathbf{b}^{\text{BR}} _{c,s}({\mathbf{a}_{c,s}^\text{BR}})^{T}}e^{\frac{-j2\pi bW {\tau^{\text{BR}} _{c,s}}}{B}}}_{\mathbf{G}_{\text{NLOS}}[b]}.
\end{align}

Similar to the BS$\to$RIS channel modeling, the frequency-domain RIS$\to$UE channel $\mathbf{H}[b]=\mathbf{H}_{\text{LOS}}[b]+\mathbf{H}_{\text{NLOS}}[b]\in \mathbb {C}^{U \times N}$ can be expressed as 
\begin{align}\label{RISUE}
\mathbf{H}[b] =&\underbrace{\sqrt{G_\text{R}G_\text{U}L_{\text{LOS}}^{{\text{RU}}}}\mathbf{u}_{\text{LOS}}^\text{RU}({\mathbf{b}}_{\text{LOS}}^\text{RU})^{T} \odot \left(\mathbf{H}^x_c \otimes \mathbf{H}^z_c\right)e^{\frac{-j2\pi bW {\tau_\text{LOS}^\text{RU}}}{B}}}_{\mathbf{H}_{\text{LOS}}[b]} \nonumber \\
& + \underbrace{{{\gamma }^\text{RU}\sum\limits_{c=1}^{{{C_\text{s}}}}{\sum\limits_{s=1}^{{{{{S}_{c}}}}}{{{{\varsigma }^{\text{RU}}_{c,s}}}}}\sqrt{G_\text{R}G_\text{U}L_{c,s}^{\text{RU}}}\mathbf{u}^\text{RU}_{c,s}(\mathbf{b}^{\text{RU}} _{c,s})^T}e^{\frac{-j2\pi b W \tau ^\text{RU}_{c,s}}{B}}}_{\mathbf{H}_{\text{NLOS}}[b]},
\end{align}
where ${\gamma^\text{RU}}$, ${{{\varsigma }}^{\text{RU}}_{c, s}}$ and $L_{c, s}^\text{RU}$ are the normalization factor, the complex gain and the path loss for scatterer $(c, s)$ in the RIS$\to$UE link, respectively. $\mathbf{u}^\text{RU}_\text{LOS}\in {\mathbb{C}^{U \times 1}}$ is the near-field ULA array response for the LOS path, while $\mathbf{u}^\text{RU}_{c,s}\in {\mathbb{C}^{U \times 1}}$ denotes the far-field array response for scatterer $(c,s)$. $\mathbf{b}^\text{RU}_{c,s} =\mathbf{b}_{x}(f_b,\phi^\text{RU}_{c,s}, \varphi^\text{RU}_\text{c,s}, r^\text{RU}_{c,s}) \otimes \mathbf{b}_{z}(f_b,\varphi^\text{RU}_{c,s}, r^\text{RU}_{c,s})\in {\mathbb{C}^{N \times 1}}$ denotes the receiving response at the RIS. Parameters $\phi^\text{RU}_{c,s}$  ($\phi^\text{RU}_\text{LOS}$) and $\varphi^\text{RU}_{c,s}$ ($\varphi^\text{RU}_\text{LOS}$) denote the azimuth and elevation of AoD of scatterer path $(c,s)$ (LOS path) at the RIS, respectively. 

\subsection{Wideband Beamforming Architecture in RIS Systems}
{Compared to conventional ELAA systems, the challenges of near-field beam focusing and the wideband beam split effect in RIS communications are significantly heightened. Firstly, in extremely-large RIS systems, the cascaded channel model in the BS$\to$RIS$\to$UE link comprises two near-field channel components, creating a double-hop beam focusing characteristic. Additionally, similar to the analog beamformer in the hybrid beamforming architecture at the BS, the beams at different subcarrier frequencies generated by the frequency-independent phase shifting circuit at the RIS point to different physical directions, resulting in a specific near-field double beam split effect for the considered RIS system. Finally, the active hybrid beamforming at the BS and the passive beamfroming at the RIS is highly coupled, necessitating joint design to optimize the overall communication system performance. Consequently, near-field wideband beamforming in RIS systems presents greater challenges than in conventional ELAA systems.}

\subsubsection{TTD-Based Hybrid Beamforming at the BS} To exhibit the {frequency-dependent} hybrid beamforming at the BS, the TTD-based hybrid beamforming architecture is adopted \cite{Dai2022Delay}. Let $\mathbf{F}_\text{PS}=\operatorname{blkdiag}\left(\mathbf{F}_{\text{PS},1}, \ldots, \mathbf{F}_{\text{PS},M_\text{RF}}\right) \in \mathbb{C}^{M \times K{M_\text{RF}}}$ denote the analog beamformer achieved by PSs, where $\mathbf{F}_{\mathrm{PS}, n}=\operatorname{blkdiag}\left(\mathbf{f}_{m_\text{RF}, 1}, \ldots, \mathbf{f}_{m_\text{RF}, K}\right) \in \mathbb{C}^{M \times K} (1\le m_\text{RF} \le {M_\text{RF}})$ denotes the PS-based analog beamformer for the subarray connected to the $m_\text{RF}$-th RF chain, and $\mathbf{f}_{m_\text{RF},k} \in \mathbb{C}^{P\times 1} (1\le k \le K)$ denotes the analog beamformer connected to the $k$-th TTD of this chain. {$\mathbf{F}_{\mathrm{BB},b}\in \mathbb{C}^{M_\text{RF} \times N_\text{s}}$} denotes the digital beamformer at subcarrier $b$, where $N_\text{s}$ denotes the number of transmitting data stream. According to \cite{Dai2022Delay}, the TTD-based analog beamformer at subcarrier $b$ is given by
\begin{align}\label{TTD}
\mathbf{T}_{b}=\operatorname{blkdiag}\left(e^{-j 2 \pi f_{b} \mathbf{t}_{1}}, \ldots, e^{-j 2 \pi f_{b} \mathbf{t}_{M_{\mathrm{RF}}}}\right),
\end{align}
where $\mathbf{t}_{m_\text{RF}}=[t_{m_\text{RF},1}, \ldots, t_{m_\text{RF},K}] \in \mathbb{R}^{K \times 1}$ denotes the time-delay vector realized by the TTDs connected to the $m_\text{RF}$-th RF chain. The time delay of each TTD needs to satisfy the maximum delay constraint, i.e., $t_{m_\text{RF},k} \in [0, t_\text{max}], \forall m_\text{RF} = 1, \ldots, M_\text{RF}, \forall k= 1, \ldots, K$.

Hence, the transmitted signal ${\mathbf{x}}_{b} \in \mathbb{C}^{M \times 1}$ at the BS at subcarrier $b$ is given by
\begin{align}\label{tSig}
{\mathbf{x}}_{b}=\mathbf{F}_{\mathrm{PS}} \mathbf{T}_{b} \mathbf{F}_{\mathrm{BB},b} {\mathbf{s}}_{b},
\end{align}
where $\mathbf{s}_b \in \mathbb{C}^{N_\text{s} \times 1}$ denotes the information symbols at subcarrier $b$, satisfying $\mathbb{E}[\mathbf{s}_b \mathbf{s}^H_b] = \frac{1}{N_\text{s}}\mathbf{I}_{N_\text{s}}$. 

\subsubsection{TTD-Based Phase-Shifting at the RIS} For the classic RIS architecture, the refection coefficients can be expressed as ${\bm {\theta}}  = {[{\beta _1}{e^{j{\theta _1}}},{\beta _2}{e^{j{\theta _2}}}, \ldots ,{\beta _N}{e^{j{\theta _N}}}]^T} \in {\mathbb{C}^{N \times 1}}$, where ${\theta }_{i}(i=1,\ldots,N)$ and ${\beta _i} \in \{ 0,1\} $ denote the phase shift and the ON/OFF state \textcolor{black}{of the} $i$-th RIS element, respectively. {To mitigate the beam split effect at the RIS, as discussed in \cite{Hao2023The, an2021reconfigurable} and \cite{10263619}, the frequency-dependent RIS architecture with fully-connected TTD units is employed, in which each RIS element is connected to an independent TTD unit. However, the power consumption and hardware cost associated with TTD units are significantly higher than those of conventional PSs. Inspired by the sub-connected TTD architecture in \cite{Su2023Wideband}, we develop a sub-connected TDD-RIS architecture, as depicted in Fig.~\ref{fig1}(a), to realize the energy-efficiency frequency-dependent phase shifting operation, in which each subarray at the RIS is connected to a common TTD unit.} 

The TTD-RIS is divided into $S$ = $S_1$ × $S_2$ subarrays, where each subarray consists of $\bar{S}$ = $\bar{S}_1$ × $\bar{S}_2$ elements, i.e., $\bar{S}_1=\frac{N_1}{S_1}$ and $\bar{S}_2=\frac{N_2}{S_2}$. To deal with the beam split effect at the RIS, each element of the subarray is equipped with double PS layers and is connected to a common TTD module. The received signal at each RIS element first passes through the first-layer PS with reflection coefficients ${\bm {\theta}_1} \in {\mathbb{C}^{N \times 1}}$, which aims to create the constructive received signal superposition at the subarray. Then, the impinging signal at the $s$-th RIS subarray $(1\le s \le S)$ is adjusted by the common TTD module with the time delay $\nu_{s} \in [0, t_\text{max}]$ to realize the {frequency-dependent} phase shifting. Finally, the signal passes through the second-layer PS with reflection coefficients ${\bm {\theta}_2} \in {\mathbb{C}^{N \times 1}}$ to accomplish the passive beamforming at the RIS\textcolor{blue}{\footnote{{In the proposed TDD-RIS architecture with sub-connected TTD units, the conventional single-layer PS circuit used in the costly fully-connected TTD architecture needs to be extended to the double-layer PS circuit. Employing a single-layer PS in the TDD-RIS architecture could lead to incoherent mixing of the received signals at a RIS subarray after they pass through a common TTD unit. This configuration might introduce critical interference and result in severe attenuation of the desired received signal \cite{Su2023Wideband}.}}}. Hence, the equivalent refection phase shifting matrix at subcarrier $b$ for the TTD-RIS can be expressed as
\begin{align}\label{TTD-RIS}
\bar{\bm {\Theta}}_b=\bm {\Theta}_1  {\bm \Lambda}_{b} \bm {\Theta}_2,
\end{align}
where $\bm {\Theta}_1=\mathrm{diag}({\bm {\theta}_1})$, $\bm {\Theta}_2=\mathrm{diag}({\bm {\theta}_2})$, and ${\bm \Lambda}_{b}=\mathrm{diag}({\bm \Lambda}^T_{b,1},\ldots,{\bm \Lambda}^T_{b,S})$. The time delay vector ${\bm \Lambda}_{b,s}\in {\mathbb{C}^{\bar{S} \times 1}}$ at the $s$-th subarray is given by ${\bm \Lambda}_{b,s}= \mathbf{1}_{\bar{S}} \otimes e^{-j 2 \pi f_{b} {\nu}_{s}}$.

\subsubsection{Virtual Subarray-Based Phase-Shifting at the RIS} In the above TTD-RIS architecture, the additional TTD units and double-layer phase shifting circuit are required, which increases the hardware cost and energy consumption in RIS systems. {In Fig.~\ref{fig1}(b), we present an SA-RIS architecture by dividing the RIS into $B$ virtual subarrays,} in which the reflection coefficients $\tilde{\bm {\theta}}_b \in {\mathbb{C}^{\frac{N}{B} \times 1}}$ at subarray $b$ is optimized according to the channels $\mathbf{D}[b], \mathbf{G}[b]$ and $\mathbf{H}[b]$ at subcarrier $b$. In this case, the refection phase shifting matrix at the SA-RIS is given by $\tilde{\mathbf{\Theta}}=\mathrm{diag}(\tilde{\bm{\theta}}^T_{1},\ldots,\tilde{\bm{\theta}}^T_{B})$. 

\textbf{\emph{Remark 1:}} On the one hand, compared to the TTD-RIS architecture, the SA-RIS architecture does not require the additional hardware cost, while the beamforming gain will be reduced due to the RIS aperture shrinkage at the specific subcarrier. On other hand, for the specific design of optimization algorithms, the frequency-dependent phase shifting composed of $\bm {\Theta}_1$, $\bm {\Theta}_2$ and $\bm{\Lambda}_{b}$ at the TTD-RIS architecture is more complex, while the dimension of phase shifting at the SA-RIS architecture is similar to the classic RIS architecture. 

\subsection{Problem Formulation}
In the downlink signal transmission at the $q$-th slot, the received signal ${\mathbf{y}_{q,b}} \in {\mathbb{C}^{U \times 1}}$ of UE at subcarrier $b$ can be expressed as
\begin{align}\label{8eq}
{\mathbf{y}_{q,b}}=\sqrt{P_t} \left( {\mathbf{{H}}{_b}{\bm {\Theta}^f_{q,b}}\mathbf{G}_b+\mathbf{{D}}{_b}} \right) {\mathbf{x}}_{q,b}+{\mathbf{n}_{q,b}},
\end{align}
where $P_t$ is the transmission power per stream and ${{\mathbf{n}}_{q,b}}\sim\mathcal{C}\mathcal{N}(0,{{{\sigma}}_b^{2}{\mathbf{I}}_U})$ denotes complex Gaussian noise. According to different RIS architectures, ${\bm {\Theta}^f_{q,b}} =\mathrm{diag}({\bm {\theta}^f_{q,b,1}},\ldots, {\bm {\theta}^f_{q,b,N}})$, $\forall f\in \{\mathbb{P}, \mathbb{T},\mathbb{V}\}$, denotes the corresponding RIS reflection coefficients at slot $q$, in which the indicator symbol $\mathbb{P}$, $\mathbb{T}$, and $\mathbb{V}$ are associated with the frequency-independent RIS, the TTD-RIS, and the SA-RIS, respectively. 

In this work, we aim to maximize the spectral efficiency of near-field wideband RIS systems by optimizing the TTD-based hybrid beamforming at the BS and the frequency-dependent phase shifting at the RIS. Considering the widely used block fading channel assumption in RIS systems \cite{zheng2022survey}, the channels remain constant within each channel coherent block that consists of $Q$ symbol durations. The achievable communication rate of UE at subcarrier $b$ within each channel coherent block is given by 
\begin{align}\label{SR}
R_{b}=\log _{2} \mathrm{det} \left({\bf{I}}_U+\frac{P_t}{N_\text{s}\delta^2}  \mathbf{Z}_b \mathbf{A}_b  \mathbf{A}^H_b \mathbf{Z}^H_b \right),
\end{align}
where $\mathbf{Z}_b = {\mathbf{{H}}{_b}{\bm {\Theta}^f_{q}}\mathbf{G}_b+\mathbf{{D}}{_b}} $ and $\mathbf{A}_b=\mathbf{F}_{\mathrm{PS}} \mathbf{T}_{b} \mathbf{F}_{\mathrm{BB},b}$. 

In the existing near-field wideband beamforming schemes in RIS systems \cite{Cheng2023Achievable, Hao2023The}, the CSI is assumed to be perfectly known when optimizing the involved beamforming variables. In fact, for near-field wideband RIS systems, the accurate channel acquisition is challenging due to the extremely large-scale antenna arrays and the passive characteristic of RIS. Considering the typical division duplex systems with the channel reciprocity, the downlink channel can be obtained by estimating the uplink channel. Hence, the effective spectral efficiency can be expressed as
\begin{align}\label{ESR}
R = \frac{Q-Q_\text{tr}}{Q(L_\text{CP}+B)}\sum\limits_{b=1}^B R_{b},
\end{align}
where $Q_\text{tr} \le Q$ denotes the number of symbol durations for the channel training, and $L_\text{CP}$ denotes the length of cyclic prefix in OFDM systems.

\textbf{\emph{Remark 2:}} In the considered near-field wideband RIS systems, $BMU(N+1)$ unknown entries need to be estimated in the channel estimation, which requires the high pilot training overhead $Q_\text{tr}$. In spite of the fact that several low-overhead channel estimation schemes have been investigated for far-field or narrowband RIS systems\cite{zheng2022survey, 9500188}, e.g., the compressed sensing approach by exploiting the channel sparsity and the deep learning-based intelligent channel estimation scheme \cite{10313112}, the aforementioned channel estimation approaches are hard to directly extended to near-field wideband RIS systems due to the specific near-field radiation and beam split effect. Consequently, the optimization of effective spectral efficiency not only depends on the beamforming design, but is also related to the channel estimation scheme.


To sum up, the spectral efficiency maximization problem in near-field wideband RIS systems can be formulated as
\begin{subequations}\label{SR}
\begin{align}
\max _{\mathbf{F}_{\mathrm{PS}}, \mathbf{T}_{b}, \mathbf{F}_{\mathrm{BB}, b},\hfill\atop
\scriptstyle {\bm {\Theta}^f_{b}}, Q_\text{tr}} & R\left(\mathbf{F}_{\mathrm{PS}}, \mathbf{T}_{b}, \mathbf{F}_{\mathrm{BB}, b}, {\bm {\Theta}^f_{b}}, Q_\text{tr}\right) \label{Za} \\
\text { s.t. } & \left\|\mathbf{F}_{\mathrm{PS}} \mathbf{T}_{b} \mathbf{F}_{\mathrm{BB}, b}\right\|_{F}^{2} = \rho, \forall b, \label{Zb}\\
& |\mathbf{f}_{m_\text{RF},k}| =1, \forall m_\text{RF}, \forall k , \label{Zc}\\
& |[{\bm {\theta}^f_{b}}]_{i}| =1, i =\{1,\ldots,N\}, \label{Zd}\\  
& \mathbf{T}_{b} \in \mathcal{T}_{b}, \forall b, \label{Ze}
\end{align}
\end{subequations}
where $\rho$ denotes the transmit power available of the precoder for each subcarrier at the BS. $\mathcal{T}_{b}$ is a feasible set of the TTD-based analog beamformers imposed by the structure in \eqref{TTD} and the maximum time delay constraint. Note that the frequency-dependent phase shifting ${\bm {\Theta}^\mathbb{T}_{b}}$ at the TTD-RIS architecture involves three coupled variables, i.e., $\bm {\Theta}_1$, $\bm {\Theta}_2$ and $\bm{\Lambda}_{b}$, to be optimized, while the optimization of ${\bm {\Theta}^\mathbb{V}_{b}}$ is composed of $B$ independent phase shifting subarrays for the SA-RIS architecture. 

\section{Deep Learning Based End-to-End Beamforming Framework}
\begin{figure*}[t]
	\centerline{\includegraphics[width=6.0in]{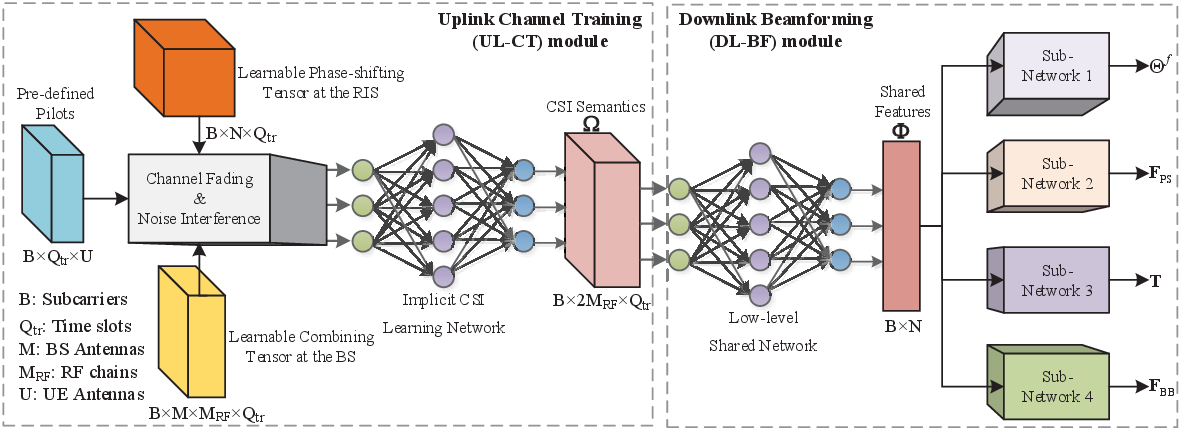}}
	 \caption{The proposed E2E beamforming framework in near-field wideband RIS systems.}
	\label{Framework}
\end{figure*}

{To solve the non-convex and high-dimensional optimization problem in \eqref{SR}, we develop an E2E beamforming optimization framework by leveraging the powerful non-linear mapping ability of the deep learning model}. As illustrated in Fig.~\ref{Framework}, the overall E2E framework can be divided into the UL-CT module in the uplink pilot transmission stage and the DL-BF module in the downlink signal transmission.

\subsection{Deep Learning-Based Uplink Channel Training}
For the design of channel estimation scheme in RIS systems, the channel estimation performance is related to the joint design of the channel estimator and the RIS reflection protocol. For instance, in \cite{9053695}, the DFT-based reflection protocol has been proven to be optimal for the classic minimum variance unbiased estimator in narrowband far-field RIS systems. However, the existing reflection protocol need to be further developed for near-field wideband RIS systems. Firstly, the assumption of planar wavefront in far-field communications is no longer applicable for near-field systems, while the RIS reflection protocol need to match the spherical wavefront characteristics instead of the far-field assumption of planar wavefront. Secondly, in the reflection protocol design for wideband RIS systems, the frequency-dependent hybrid beamforming and phase shifting involve the new time-delay dimension. Thirdly, the key characteristic of deep learning estimator is the adaptive learning ability for the latent representation of the wireless signal, while the advantage of RIS is to operate the wireless environment as a passive reflector. However, for the deep learning-based channel estimator, the reflection protocol design in near-field wideband RIS systems has not been investigated. 

Consequently, in the proposed E2E optimization framework, we develop a learnable UL-CT module to learn high-dimensional channel semantics, in which the trainable wideband phase shifting at the RIS and the combining matrix at the BS are designed. In contrast to the pre-defined RIS reflection pattern in the existing channel estimation works \cite{9053695, zheng2022survey}, the phase shifting and combining matrix in the proposed E2E model can be adaptively tuned according to dynamic wireless environments. Suppose $Q_\text{tr}$ OFDM pilots are used for the channel training, the received pilot signal $\mathbf{Y}_{q,b} \in {\mathbb{C}^{M_\text{RF} \times 1}}$ at slot $q(1 \le q \le Q_\text{tr})$ in subcarrier $b$ of the BS is given by
\begin{align}\label{pilot}
{\mathbf{Y}_{q,b}}= {\mathbf{W}}_{q,b} \left( {\mathbf{{H}}_b{\bm {\Theta}^f_b}\mathbf{G}_{q,b}+\mathbf{{D}}_b} \right) {\mathbf{X}}^\text{tr}_{q,b}+{\mathbf{n}^\text{tr}_{q,b}},
\end{align}
where $\mathbf{X}^\text{tr}_{q,b} \in {\mathbb{C}^{U \times 1}}$ denotes the pilot signal sent by the UE at slot $q$ in subcarrier $b$, $\mathbf{W}_{q,b} \in {\mathbb{C}^{M_\text{RF} \times M}}$ denotes the uplink combining matrix in subcarrier $b$ at the BS, and complex Gaussian noise follows ${{\mathbf{n}}^\text{tr}_{q,b}}\sim\mathcal{C}\mathcal{N}(0,{{{\sigma}}_b^{2}{\mathbf{I}}_{M_\text{RF}}})$. In this work, the phase shifting ${\bm {\Theta}^f_{q,b}}$, $\forall f\in \{\mathbb{P}, \mathbb{T},\mathbb{V}\}$ and the combining matrix $\mathbf{W}_{q,b}$ are designed the trainable tensors, which are optimized by utilizing the massive training data. 

In near-field wideband RIS systems, ${\bm {\Theta}^f_{q,b}}$ and $\mathbf{W}_{q,b}$ need to be restricted by the specific constraint. Specifically, for the TTD-RIS architecture as illustrated in \eqref{TTD-RIS}, the frequency-dependent phase shifting ${\bm {\Theta}^\mathbb{T}_q}$ at slot $q$ consists of the learnable phase shifting tensor {$\widehat{\bm {\Theta}}^\mathbb{T}_{q,1}\in {\mathbb{C}^{N \times N}}$, $\widehat{\bm {\Theta}}^\mathbb{T}_{q,2}\in {\mathbb{C}^{N \times N}}$, }and time delay tensor $\widehat{\mathbf{\Lambda}}_{q}\in {\mathbb{C}^{S \times B}}$. The diagonal elements in $\widehat{\bm {\Theta}}^\mathbb{T}_{q,1}$ and $\widehat{\bm {\Theta}}^\mathbb{T}_{q,2}$ should satisfy the unit modular constraint. Thus, the complex exponent function $\mathrm {exp}(\mathrm j \cdot)$ is applied to obtain the desired uplink $\widehat{\bm {\Theta}}^\text{up}_{q,1}$ and $\widehat{\bm {\Theta}}^\text{up}_{q,2}$, which is given by
\begin{align}\label{exp1}
\left\{ \begin{array}{l}
\widehat{\bm {\Theta}}^{\text{up},\mathbb{T}}_{q,1}= \mathrm {exp}\left({\mathrm j \cdot \widehat{\bm {\Theta}}^\mathbb{T}_{q,1}}\right),\\
\widehat{\bm {\Theta}}^{\text{up},\mathbb{T}}_{q,2}= \mathrm {exp}\left({\mathrm j  \cdot \widehat{\bm {\Theta}}^\mathbb{T}_{q,2}}\right), \forall q.\\
\end{array} \right.
\end{align}

Due to the maximum time-delay constraint in the TTD unit, each learnable time delay element $\widehat {\nu}_{s,b}$ in $\widehat{\mathbf{\Lambda}}_{q}$ need to be normalized by the maximum delay $t_\text{max}$, which is given by
\begin{align}\label{time}
\widehat{\nu}^\text{up}_{s,b}= t_\text{max} \frac{1}{1+e^{-\widehat {\nu}_{s,b}}}, \forall s,b.
\end{align}

In the SA-RIS without TTD units, the uplink phase shifting $\widehat{\bm {\Theta}}^{\text{up}, \mathbb{V}}_q$ is directly normalized by utilizing the complex exponent function. For the learnable uplink combining tensor at the BS, the uplink ${\mathbf{W}}^\text{up}_{q,b}$ should satisfy the constant modular constraint, which is given by \cite{Wu2023Deep}
\begin{align}\label{upcom}
{\mathbf{W}}^\text{up}_{q,b}= \frac{1}{\sqrt{M}}\mathrm {exp}\left({1\mathrm j \cdot {\mathbf{W}}_{q,b}}\right), \forall q,b.
\end{align}

After $Q_\text{tr}$ pilots transmission, we can obtain $B \times M_\text{RF}   \times Q_\text{tr}$ observation tensor ${\mathbf{Y}}^\text{P}$. To facilitate data processing in the neural network, the complex-to-real operation is used to separate the real and imaginary parts of ${\mathbf{Y}}^\text{P}$, and then are stacked along the antenna dimension to obtain the real-value input tensor 
${\bar{\mathbf{Y}}^\text{P}}=\{\Re({{\mathbf{Y}}^\text{P}}),{\Im}({{\mathbf{Y}}^\text{P}})\} \in \mathbb{C}^{B \times 2M_\text{RF}  \times Q_\text{tr}}$. We exploit an implicit CSI learning network ${{\mathcal{F}}^\text{C}}(\cdot)$ to map the latent CSI semantic $\bf{\Omega} $ from the observation tensor $\bar{{\mathbf{Y}}}^\text{P}$, which is given by
\begin{align}\label{exp}
{\bf{\Omega}} = {{\mathcal{F}}^\text{C}}(\bm{\omega}^\text{C}, \bar{\mathbf{Y}}^\text{P}),
\end{align}
where $\bm{\omega}^\text{C}$ denotes the trainable network parameters of the implicit CSI learning network. In Section VI-A, the detailed network architecture of the proposed UL-CT module ${{\mathcal{F}}^\text{C}}(\cdot)$ will be elaborated.

\subsection{Deep Learning-Based Downlink Wideband Beamforming}
The proposed DL-BF module ${{\mathcal{F}}^\text{B}}(\cdot)$ is composed of a low-level shared network ${{\mathcal{F}}^\text{B}_\text{sh}}(\cdot)$ and $P$ sub-networks ${\mathcal{F}}^\text{B}_p(\cdot)$, $1\le p \le P$. In the pipeline of information flow, the extracted CSI semantic $\bf{\Omega} $ in the UL-CT module is delivered to the shared network at first, which generates the shared features {${\bf{\Phi}} ={{\mathcal{F}}^\text{B}_\text{sh}}(\omega^\text{B}_\text{sh}, {\bf{\Omega}}) \in \mathbb{R}^{B \times N}$}. Then, the downlink frequency-dependent phase shifting at the RIS and the hybrid beamforming at the BS is obtained from different sub-networks, respectively, which is given by
\begin{align}\label{BF}
\left\{ \begin{array}{l}
\widehat{\bm {\Theta}}^{\text{down},f}= \mathrm {exp}\left({1 \mathrm j \cdot {\mathcal{F}}^\text{B}_1\left(\bm{\omega}^\text{B}_1,{\bf{\Phi}}\right)}\right), \forall f, \\
\widehat{\mathbf {F}}^\text{down}_\text{PS}= \mathrm {exp}\left(1 \mathrm j  \cdot {\mathcal{F}}^\text{B}_2\left(\bm{\omega}^\text{B}_2,{\bf{\Phi}}\right)  \right),   \\
\widehat{\mathbf{T}}^\text{down}_{b}= \mathrm {exp}\left(1 \mathrm j  \cdot 2\pi f_b \cdot {\mathcal{F}}^\text{B}_3\left(\bm{\omega}^\text{B}_3,{\bf{\Phi}}\right)  \right), \forall b, \\
\widehat{\mathbf {F}}^\text{down}_{\text{BB},b}= \frac{\sqrt{P_t}{\mathcal{F}}^\text{B}_4\left(\bm{\omega}^\text{B}_4,{\bf{\Phi}}\right)}{\left\|  \mathbf{F}_{\mathrm{PS}} \mathbf{T}_{b} \mathbf{F}_{\mathrm{BB},b} \right\|_{F}}, \forall b, 
\end{array} \right.
\end{align}
where $\bm{\omega}^\text{B}_\text{sh}$ and $\bm{\omega}^\text{B}_p (1\le p \le 4)$ denote the trainable network parameters of the shared network and sub-network $s$, respectively. Similar to the tensor constraints in the proposed UL-CT module, the specific constraints need to be satisfied in the output of each sub-network. Firstly, the unit-modulus constraints are applied to the phase shifting $\widehat{\bm {\Theta}}^{\text{down},f}$ at the RIS and PS-based analog beamformer $\widehat{\mathbf {F}}^\text{down}_\text{PS}$ at the BS by utilizing the complex exponent function. Secondly, the time-delay vector in TTD-based analog beamformer $\widehat{\mathbf{T}}^\text{down}_{b}$ is normalized referring to the operation in \eqref{time}. Finally, the power normalization is carried out for the digital beamformer $\widehat{\mathbf {F}}^\text{down}_{\text{BB},b}$. 

\subsection{Joint Optimization of E2E Beamforming Framework}
To maximize the spectral efficiency in \eqref{SR},  the loss function $\mathcal{L}(\bm{\omega})$ is designed as the negative spectral efficiency in the network training, which is minimized by utilizing the  gradient descent methods. The spectral efficiency maximization problem can be reformulated as
\begin{subequations}\label{minSR}
\begin{align}
\min _{ \bm{\omega}} ~& \mathcal{L}(\bm{\omega}) = -R \left(\bm{\omega}, \widehat{\mathbf{F}}^\text{down}_{\mathrm{PS}}, \widehat{\mathbf{T}}^\text{down}_{b}, \widehat{\mathbf{F}}^\text{down}_{\mathrm{BB}, b}, {\widehat{\bm {\Theta}}^{\text{down},f}_{b}}, Q_\text{tr}\right) \label{Sa} \\
& \text { s.t. } \eqref{exp1}, \eqref{time}, \eqref{upcom}, \eqref{BF}
\end{align}
\end{subequations}
where ${ \bm{\omega}}$ denotes the overall network trainable parameters composed of the UL-CT module and the DL-BF module. 

In the network training, a computational-efficiency stochastic optimization with first-order gradients is adopted to update ${ \bm{\omega}}$ \cite{kingma2014adam}, which has been proven to be robust and well-suited to a wide range of non-convex optimization problems. Let $g_t={ {\nabla _{{ \bm{\omega}}}}\cal L}_t({ \bm{\omega}})=\frac{{\partial \mathcal{L}({{ \bm{\omega}}_t})}}{{{{ \bm{\omega}}_t}}}$ denote the gradient at timestep $t$ in the network training. The moving averages gradient $m_t$ and the squared gradient $v_t$ of $g_t$ at timestep $t$ are defined as
\begin{align}
m_{t} = \beta^t_{1} \cdot m_{t-1}+\left(1-\beta^t_{1}\right) \cdot g_{t},\\
v_{t}=\beta^t_{2}\cdot v_{t-1}+\left(1-\beta^t_{2}\right)\cdot g_{t}^{2},
\end{align}
where $\beta^t_{1}$ and $\beta^t_{2} \in [0,1)$ denote the hyper-parameters that control the exponential decay rates of $m_t$ and $v_t$, respectively. Furthermore, the update rule of ${ \bm{\omega}}$ at timestep $t+1$ can be expressed as
\begin{align}
{ \bm{\omega}}_{t+1} = { \bm{\omega}}_{t}-\alpha_{t} \cdot m_{t} /\left(\sqrt{v_{t}}+\hat{\epsilon}\right),
\end{align}
where $\alpha_{t}=\alpha \cdot \sqrt{1-\beta_{2}^{t}} /\left(1-\beta_{1}^{t}\right)$ denotes the adaptive learning rate at timestep $t$ based on the default learning rate $\alpha$, and $\hat{\epsilon}$ is a {regularization term to avoid dividing by zero.} 

\textbf{\emph{Remark 3}}: Due to the low computational complexity in the weight update of network, the above stochastic optimization principle in the proposed E2E  beamforming framework is a versatile algorithm that scales to large-scale high-dimensional non-linear mapping problems. In addition, this stochastic optimization can adaptively adjust the learning rate in the network training process, thereby improving the convergence speed and generalization ability of the beamforming model for dynamic wireless environments. 

\section{Signal-Guided Beamforming Network {Architecture} in Near-Field Wideband RIS Systems}
In this section, we will present the specific beamforming network components based on the proposed E2E beamforming framework in Section III, which consists of the polar-attention network {architecture} in the UL-CT module and multi-task network {architecture} in the DL-BF module.

\subsection{Polarized Self-Attention for Channel Semantic Learning}
\begin{figure}[t]
	\centerline{\includegraphics[width=3.2in]{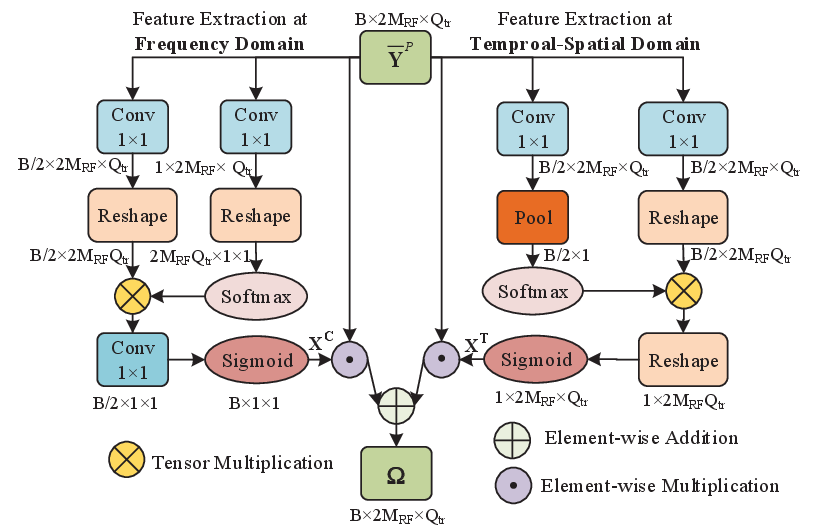}}
	\caption{{Polarized self-attention mechanism for channel semantic learning in the UL-CT module.}}
	\label{ploarAtten}
\end{figure}

In the proposed E2E models, the UL-CT module need to learn the efficient channel semantics from the received pilot signal, which facilitates the beamforming optimization in the subsequent DL-BF module. {In the popular network architectures for deep learning enabled MIMO communications, the classical convolutional neural network (CNN) with spatial modeling ability is usually used as the basic network backbone \cite{9220182}. However, due to the limited receptive field of local convolution window, the insufficient ability to extract global information of CNN has been widely investigated.} Accordingly, the promising self-attention mechanism has drawn enthusiastic concern. Motivated by the extensive representation learning ability of the self-attention mechanism, we exploit a dedicated polarized self-attention (PSA) mechanism to characterize the implicit CSI semantics, which leverages the specific physical knowledge of wireless communications data.

In contrast to the dataset in computer vision or nature language models, each dimension of input tensor ${\bar{\mathbf{Y}}^{P}}$ in the proposed UL-CT module has the specific physical implications, which represents time, frequency, and antenna domains, respectively. In the vanilla self-attention mechanism, the attention operation is only carried out in the spatial domain of input tensor, i.e., the dimension $2M_\text{RF}  \times Q_\text{tr}$ in ${\bar{\mathbf{Y}}^{P}}$. In this work, the self-attention mechanism is introduced {into} the frequency and temporal-spatial domain of ${\bar{\mathbf{Y}}^{P}}$ by designing the dedicated PSA module \cite{liu2021polarized}, respectively. In Fig.~\ref{ploarAtten}, we present the detailed operation of the PSA module, which consists of the feature extraction branches in the frequency and time-spatial domain. Specifically, in the frequency feature extraction branch, the output features ${{\mathbf{X}}^\text{F}} \in \mathcal{R}^{B \times 1 \times 1}$ can be expressed as
\begin{align}
{{\mathbf{X}}^\text{F}}=\text{Sig}\left(\mathbf{W}^\text{F}_{z}(\text{Re}(\mathbf{W}^\text{F}_{v}({\bar{\mathbf{Y}}^{P}})) \times \text{Soft}(\text{Re}(\mathbf{W}^\text{F}_{q}({\bar{\mathbf{Y}}^{P}})))\right),
\end{align}
where $\mathbf{W}_{i}^\text{F}, i=\{q,v,z\}$ denotes $1\times1$ convolutional layer to reduce or increase the frequency-domain dimension of ${\bar{\mathbf{Y}}^{P}}$. {Specifically, $\mathbf{W}^\text{F}_{q}$, $\mathbf{W}^\text{F}_{v}$, and $\mathbf{W}^\text{F}_{z}$ are composed of $1$, $B/2$, and $B$ convolutional filters, respectively.} Function $\text{Re}(\cdot)$ represents the tensor reshape operator to adjust the dimension of different feature tensors. Functions $\text{Soft}(\cdot)$ and $\text{Sig}(\cdot)$ denote Softmax and Sigmoid activation function, respectively, which can be expressed as
\begin{align}
\text{Soft}(\mathbf{x}) =  \frac{e^{\mathbf{x}_i}}{\sum_{i = 1}^{L}{e^{\mathbf{x}_i}}},\\
\text{Sig}(\mathbf{x}) = \frac{1}{1+e^{- {\mathbf{x}}_{i}}},
\end{align}
where $\mathbf{x} \in \mathcal{R}^{L \times 1}$ denotes a feature tensor. 

Similarly, the output features in the time-spatial feature extraction branch ${{\mathbf{X}}^\text{T}} \in \mathcal{R}^{1 \times 2M_\text{RF}  \times Q_\text{tr}}$ can be expressed as
\begin{align}
{{{\mathbf{X}}^\text{T}}}=\text{Sig}\left(\text{Re}(\mathbf{W}^\text{T}_{v}({\bar{\mathbf{Y}}^{P}})) \times \text{Soft}(\mathcal{G}(\mathbf{W}^\text{T}_{q}({\bar{\mathbf{Y}}^{P}}))\right),
\end{align}
where {$\mathbf{W}_i^\text{T}, i=\{q,v\}$ denotes a $1\times1$ convolutional layer composed of $B/2$ convolutional filters, and $\mathcal{G}(\cdot)$ denotes the global average pooling operator.} For the feature tensor $\mathbf{F}^\text{T} = \mathbf{W}_{q}({\bar{\mathbf{Y}}^{P}}) \in \mathcal{R}^{B/2 \times 2M_\text{RF}  \times Q_\text{tr}}$ obtained by a convolutional layer with the $1\times 1$ kernel, the feature vector ${\mathbf{z}} = \left[ {{z_1}, \cdots, {z_b}, \cdots, {z_{{B/2}}}} \right] \in {\mathbb{R}^{{B/2}\times 1}}$ after pooling operation $\mathcal{G}(\mathbf{F}^\text{T})$ is given by
\begin{align}
{z_b} = \mathcal{G}(\mathbf{F}^\text{T}) = \frac{1}{{2M_\text{RF}  \times Q_\text{tr}}}\sum\limits_{{m} = 1}^{2M_\text{RF} } {\sum\limits_{{q} = 1}^{Q_\text{tr}} {\mathbf{F}^\text{T}_{b}} } ({m},{q}).
\end{align}

Finally, the frequency and time-spatial features are fused in the UL-CT module, which can be expressed as 
\begin{align}
\bf{\Omega} = {{\mathbf{X}}^\text{F}} \odot^\text{F} {\bar{\mathbf{Y}}^{P}} + {{\mathbf{X}}^\text{T}} \odot^\text{T} {\bar{\mathbf{Y}}^{P}},
\end{align}
where $\odot^i,i\in \{\text{F}, \text{T}\}$ denotes a channel-wise or spatial-wise multiplication operator, respectively. Note that compared to the classic self-attention mechanism, the PSA module has lower computational complexity due to the separable dual branch attention architecture\textcolor{blue}{\footnote{{For the observation pilot tensor $\bar{{\mathbf{Y}}}^\text{P} \in \mathbb{R}^{B \times M_\text{RF}  \times Q_\text{tr}}$ in the UL-CT module, the uplink pilots in each subcarrier are independently designed, and hence the frequency domain features are processed in an independent branch. Since the temporal-spatial domain of the received pilot signal is strongly correlated and the number of RF chains $M_\text{RF}$ is relatively small, the temporal-spatial domain data tensor is simultaneously operated in the UL-CT module. This arrangement efficiently facilitates the feature learning process for $\bar{{\mathbf{Y}}}$.}}}.

\subsection{Signal-Guided Network for Beamforming Design}
\begin{figure}[t]
	\centerline{\includegraphics[width=3.0in]{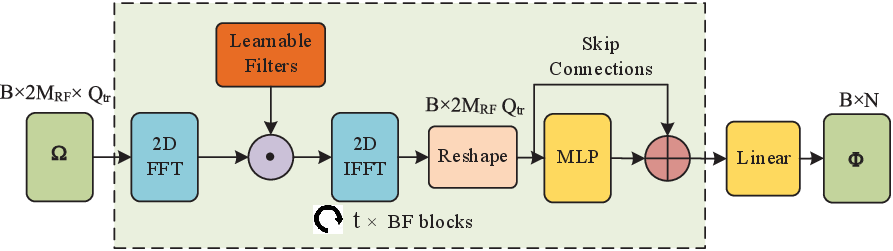}}
	{{\caption{{Signal-guided shared network architecture for beamforming design in the DL-BF module.}}}}
	\label{FFTAtten1}
\end{figure}

The DL-BF network consists of the low-level shared network to extract the shared embedded features and multiple independent sub-networks to jointly design near-field wideband beamforming.

\subsubsection{Signal-Guided Shared Network Architecture}
In the large-scale array communications at high frequencies, wireless channels present the natural sparsity due to limited scatterer paths. Considering the severe energy spreading effect, the sparsity representation in the far-field channel, such as the DFT transform-based angular domain sparsity \cite{Gao2022Data}, may no longer be applicable to the near-field channel. However, as a classic digital signal processing tool, the DFT principle is still useful for guiding the network architecture design \cite{rao2021global}, especially for the dedicated network in wireless communications. In the proposed DL-BF module, we incorporate Fourier transform to the conventional self-attention mechanism and employ learnable filters to interchange information globally among the feature tokens in the Fourier domain, which is termed as a signal-guided deep learning approach.

{As shown in Fig.~5, for the feature $\bf{\Omega}$ obtained by the UL-CT module in Fig.~\ref{ploarAtten}, the two-dimensional DFT {at subcarrier $b$} is carried out at first,} which can be expressed as
\begin{align}
{\bf{F}}^\text{DFT}_b[u, v]=\mathcal{F}({\bf{\Omega}}_b)=\sum_{m=0}^{2M_\text{RF}-1} \sum_{q=0}^{Q_\text{tr}-1} {\bf{\Omega}}_b[m, q] e^{-j 2 \pi\left(\frac{u m}{2M_\text{RF}}+\frac{v q}{Q_\text{tr}}\right)}.
\end{align}

Since the DFT of real input tensor ${\bf{\Omega}}_b[m, q]$ satisfies the conjugate symmetry property, i.e., ${\bf{F}}^\text{DFT}_b[2M_\text{RF}-u, Q_\text{tr}-v]=\left({\bf{F}}^\text{DFT}_b[u, v]\right)^*$, this property implies that the half of ${\bf{F}}^\text{DFT} \in {\mathbb{R}^{B \times 2M_\text{RF} \times {Q_\text{tr}/2}}}$ contains the full information about the frequency characteristics of $\bf{\Omega}$. In this case, we can take only the half of the values in ${\bf{F}}^\text{DFT}$ but preserve the full information, which can reduce the network parameters and computational cost. As mentioned above, the pure DFT is hard to fully exploit the near-field channel sparsity, we further employ a learnable filter ${\bf{\Psi}}\in {\mathbb{R}^{B \times 2M_\text{RF} \times {Q_\text{tr}/2}}}$ in the frequency domain to modulate the spectrum of ${\bf{F}}^\text{DFT}$, which can be expressed as
\begin{align}
\bar{\mathbf{F}}^\text{DFT}={\bf{\Psi}} \odot {\mathbf{F}}^\text{DFT}.
\end{align}

{For the practical implementation of DFT in deep neural networks, the existing efficient algorithms for computing the DFT, i.e., the well-known fast Fourier transform (FFT) algorithms, are well supported by both GPU and CPU architectures, thanks to the acceleration libraries, e.g., cuFFT and mkl-fft \cite{paszke2019pytorch}.} {The learnable filter ${\bf{\Psi}}$ is designed as a tensor with trainable parameters, which can be adaptively optimized according to the training data in the network training stage. In the test stage, the parameters of ${\bf{\Psi}}$ can be directly determined based on the received pilot tensor ${\bar{\mathbf{Y}}^\text{P}}$.}

Next, the inverse DFT is used to transform $\bar{\mathbf{F}}^\text{DFT}$ into the original domain, i.e., ${\bf{F}}^\text{IDFT}_b=\mathcal{F}^{-1}(\bar{\mathbf{F}}^\text{DFT}_b) \in {\mathbb{R}^{B \times 2M_\text{RF} \times {Q_\text{tr}}}},\forall b$. Then, ${\bf{F}}^\text{IDFT}$ is flatten to the feature tensor $\mathbf{F}^\text{M} \in {\mathbb{R}^{2M_\text{RF} Q_\text{tr} \times B}} $ along the time-spatial dimension. A multi-layer perceptron (MLP) block that is composed of two linear layers, are used to realize the frequency-domain feature interaction of $\mathbf{F}^\text{M}$, which is given by 
\begin{align}
{\mathbf{A}^\text{f}= \rm{GeLU}(\mathbf{\mathbf{F}^\text{M}}\mathbf{W}_1)\cdot \mathbf{W}_2+\mathbf{\mathbf{F}^\text{M}},}
\end{align}
where the first linear layer $\mathbf{W}_1  \in {\mathbb{R}^{B \times \upsilon B}}, (\upsilon \ge 1)$ projects the feature $\mathbf{F}^\text{M}$ into the high-dimension representation space. The second linear layer $\mathbf{W}_2  \in {\mathbb{R}^{\upsilon B \times B}}$ is used to recover the desired channel dimension again. Function $\rm{GeLU}(\cdot)$ denotes the Gaussian error linear unit activation function to provide the non-linearity of feature transformation. 

{In the signal-guided shared network architecture, $t$ learnable DFT blocks are stacked to extract the latent representation of CSI semantics in the Fourier domain.} Finally, $\mathbf{A}^\text{f}$ is converted into a feature tensor ${\bf{\Phi}} \in \mathbb{R}^{B \times N}$ by a linear layer, which is used as the input tensor for subsequent beamforming sub-networks.


\subsubsection{High-Level Sub-Network Architecture}
Considering the network complexity and the convenience of tensor operation in different sub-networks, the stacked linear layers are used as the basic component of sub-network architecture. 
{Specifically, an MLP-Mixer module in \cite{tolstikhin2021mlp} that consists of two MLP blocks, is designed to refine the feature extraction of the shared feature tensor ${\bf{\Phi}}$. In the MLP-Mixer module, the first MLP block acts on columns of ${\bf{\Phi}}$ to obtain the feature tensor ${\bf{\Phi}}_1$, while the second MLP block in the MLP-Mixer module acts on columns of ${\bf{\Phi}}_1$ to obtain the output feature tensor ${\bf{\Phi}}_2$. By leveraging the MLP operations along the columns and rows, the cross-variate information with the global dependency can be extracted.}
 
{In the output layers of  sub-networks, the linear layers are designed to realize the dimension alignment between the feature tensor ${\bf{\Phi}}_2$ and the desired frequency-dependent phase shifting, as well as the hybrid precoding matrices.} For the frequency-dependent phase shifting at the TTD-RIS architecture, three parallel linear layers are used to construct the output $\bm {\Theta}^\text{down}_1$, $\bm {\Theta}^\text{down}_2$ and $\bm{\Lambda}^\text{down}_{b}$, and then are normalized according to \eqref{exp1} and \eqref {time}, respectively. The frequency-dependent phase shifting ${\bm {\Theta}^\mathbb{T}_{b}}$ is obtained by aggregating $\bm {\Theta}^\text{down}_1$, $\bm {\Theta}^\text{down}_2$ and $\bar{\mathbf{T}}^\text{down}_{b}$ in \eqref{TTD-RIS}. In the SA-RIS architecture, we utilize a linear layer with $N$ neurons to obtain ${\bm {\Theta}^\mathbb{V}}$, in which $N$ neurons are divided into $B$ groups and each group consists of $N/B$ neurons to map the phase shifting of a RIS subarray. Similarly, the frequency-dependent hybrid precoding matrices at the BS composed of the analog beamformer $\mathbf{F}^\text{down}_{\mathrm{PS}}$, the time-delay vector $\mathbf{T}^\text{down}_{b}$ and the digital beamformer $\mathbf{F}^\text{down}_{\mathrm{BB},b}$, are obtained by constructing three sub-networks with the given constraints in \eqref{BF}. 

\subsection{Model Deployment and Complexity Analysis}

{The model deployment of the proposed E2E beamforming architecture can be divided into three stages: offline training, online finetuning, and real-time testing stages. 1) In the offline training stage, given a general communication dataset, the E2E model is trained according to the proposed optimization framework in Section III; 2) In the finetuning stage, the network parameters of the UL-CT module and the DL-BF module, the uplink wideband phase shifting at the RIS, and the combining matrices at the BS are updated according to the received pilot signal in the specific communication scenario; 3) In the testing phase, the server transmits the trained model to the BS, which can generate the desired phase shifting at the RIS and the hybrid precoding matrices at the BS for the target communication scenario.}

{In the proposed E2E beamforming framework, the time complexity of the PSA-based UL-CE module can be expressed as ${\mathcal O}\left(BM_{\rm{RF}} Q_{\rm{tr}}(7B/4+1)\right)$. In the DL-BF module, the time complexity of the low-level shared network can be represented by ${\mathcal O}\left(t(2BM_{\rm{RF}} Q_{\rm{tr}}+2\mu B^2)+2NM_{\rm{RF}} Q_{\rm{tr}}\right)$. For the TDD-RIS architecture, the time complexity of the high-level sub-netwoks is given by ${\mathcal O}\left(12\mu (B^2+C^2)+B(2+B)+NM_{\rm{RF}}(K+N_{\rm{s}})\right)$, while the time complexity of the high-level sub-netwoks can be reduced to ${\mathcal O}\left(12\mu (B^2+C^2)+N/B+NM_{\rm{RF}}(K+N_{\rm{s}})\right)$ in the SA-RIS architecture due to the relatively simpler RIS configuration. Furthermore, the space complexity of the UL-CE module is given by ${\mathcal O}\left(B(7B/4+1)\right)$. For the DL-BF module that consists of the learnable DFT and stacked linear layers, the space complexity is approximately equivalent to the time complexity.}


\section{Numerical Results}
In this section, we first introduce the simulation setups of the formulated near-field wideband systems and training hyper-parameters of the proposed models. Then, we compare the spectral efficiency of the proposed E2E models with the existing benchmarks, and further evaluate the beamforming performance under various system setups.
\subsection{Simulation Setups}
In the simulation, we set $M = 128$, $N = 16 \times 32$, $U = M_\text{RF} = N_\text{s} = 4$, $B = 16$ and $L_\text{CP} = 4$. The carrier frequency is set to $f_c= 73$ GHz and the communication bandwidth is $W = 7 $ GHz in OFDM systems. {In the clustered scatterer channel modeling, the number of clusters in both BS$\to$RIS, RIS$\to$UE and BS$\to$UE links is set to ${C}_{\text{s}}^\text{BR}={C}_{\text{s}}^\text{RU}={C}_{\text{s}}^\text{BU}=3$, while the number of scatterers within cluster $c$ is set to ${S_c^\text{BR}}=6$, ${S_c^\text{RU}}=5$ and ${S_c^\text{BU}}=4$, respectively. For each cluster $c$, the central angle of AoA $\phi _{c}$ and AoD $\varphi _{c}$ follow the uniform distribution $\phi _{c} \sim\mathcal{U}[-\pi /2,\pi /2]$ and $\varphi _{c} \sim\mathcal{U}[-\pi /2,\pi /2]$. The corresponding angular spreads are set to ${{\sigma }_{\phi }}={{\sigma }_{\varphi }}=5^{\circ}$ for scatterer paths within cluster $c$.}
{The coordinates of BS and RIS are set to $\mathbf{c}^\text{B}=\left( {x^\text{B},{y^\text{B}},{z^\text{B}}} \right)=(0,0,5)$ m and $\mathbf{c}^\text{R}=\left( {{x^\text{R}},y^\text{R},{z^\text{R}}} \right)= (0,20,5)$ m, respectively. The coordinate of UE is randomly sampled in the 1 m height with a horizontal radius of 5 m centered on RIS.} The number of TTD units connected by each RF chain is $K=16$ at the BS, in which the maximum time delay is $t_\text{max}= 5$ nanoseconds for each TTD unit. In TTD-RIS architecture, the number of subarray is ${S} = 8 \times 8$, while the number of virtual subarray at the SA-RIS architecture is fixed as the number of subcarriers $B$. The array gain at the transmit antenna, the receive antenna and the RIS element are set to $G_\text{B} =25$ dBi, $G_\text{U} =20$ dBi \cite{Wang2023STAR}, and $G_\text{R} = 5$ dBi \cite{basar2021indoor}. In the proposed UL-CT module, the channel training overhead is set to $Q_\text{tr}=NU/8$. The received SNR at the uplink pilot transmission stage is defined as $\text{SNR}_\text{R}=\frac{\left\| {\left( {\mathbf{{H}}{_b}{\bm {\Theta}^f_{b}}\mathbf{G}_b+\mathbf{{D}}{_b}} \right)\mathbf{W}_b} \right\|_F^2}{\sigma_b^2}$ for subcarrier $b$, while the transmitted SNR  at the downlink beamforming stage is defined as $\text{SNR}_\text{T} = \frac{P_t}{\sigma_0^2}$ with $\sigma_0^2 = \sigma_b^2, \forall b$. {In each training iteration of the proposed E2E models, the uplink $\text{SNR}_\text{R}$ in UL-CT module is randomly selected from the SNR range of $[0,\ldots,20]$ dB with the interval of 5 dB to improve the robustness of the trained E2E models, while the downlink $\text{SNR}_\text{T}$ in DL-BF module is fixed as 20 dB.} In the test stage, the uplink and downlink SNRs are set to $\text{SNR}_\text{R}=10$ dB and $\text{SNR}_\text{T}=20$ dB unless other specified, respectively. In this work, we compare the proposed E2E models with the following wideband beamforming benchmarks.

$\bullet$Projected gradient descent-based precoding ({PGDP}) with fully-digital beamforming architecture at the BS\cite{9343768}: A joint optimization framework for the covariance matrix of the transmitted signal and the phase shifting of RIS elements. Based on the typical PGDP method, we construct an \emph{ideal PGDP} method to characterize the performance upper bound in the formulated near-field wideband RIS system, in which the phase shifting of RIS elements is assumed to be independently designed for each subcarrier.

$\bullet$Alternative delay-phase precoding ({ADPP}) with TTD-based hybrid beamforming architecture\cite{10541333}: A dedicated hybrid beamforming architecture to deal with the beam split effect in conventional wideband MIMO systems. 

$\bullet$Alternative manifold optimization-based precoding ({AMOP}) with the classic hybrid beamforming architecture\cite{10041805}: A joint beamforming and channel reconfiguration method for RIS-aided mmWave MIMO-OFDM systems. 

In the above beamforming benchmarks, the accurate CSI is required to achieve the efficient beamforming optimization. To comprehensively present the performance advantages of the proposed E2E models, we compare the existing beamforming benchmarks with the perfect CSI at first, wherein the ideal spectral efficiency, i.e., omitting training overhead in \eqref{ESR}, is used as the performance metric. Then, we compare the practical beamforming benchmarks with the estimated CSI, in which the parallel factor (PARAFAC) decomposition-based RIS channel estimation method is used to obtain the required CSI \cite{9361077, 9366805}. Due to the full rank condition involving the LS problem, the required minimum pilot training overhead is $U(N+1)$ in the PARAFAC-based channel estimation method. 

\subsection{{Comparison} between Different Beamforming Schemes}

\begin{figure}[t]
	\centerline{\includegraphics[width=3.0in]{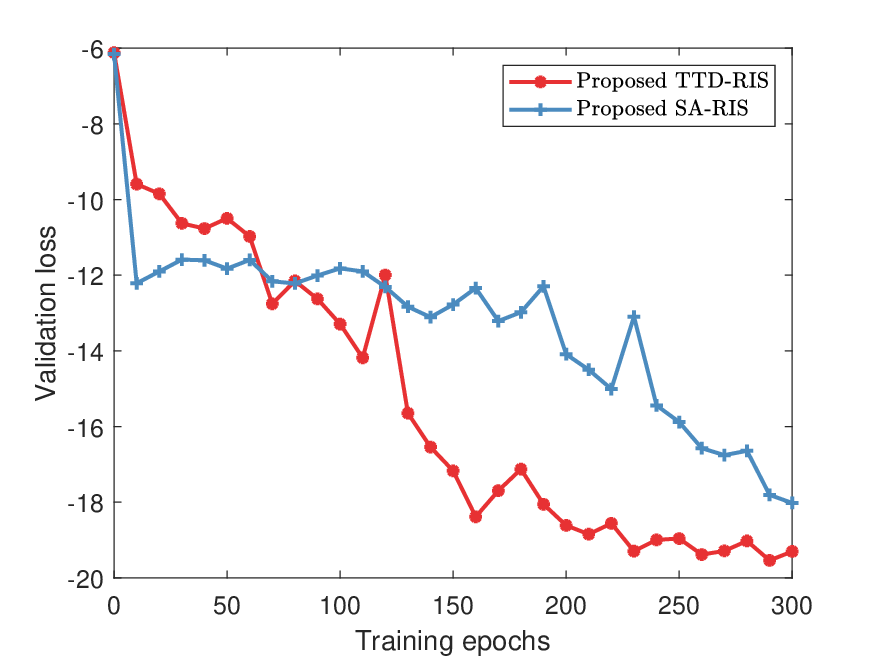}}
	\caption{{Convergence of E2E models for different RIS architectures.}}
	\label{R_Loss}
\end{figure}

{In Fig.~\ref{R_Loss}, we present the convergence of the proposed E2E models for the proposed TTD-RIS and SA-RIS architectures, in which the average loss of validation dataset in each training epoch is computed according to \eqref{minSR}. Compared to the E2E model with SA-RIS architecture, the optimization of ${\bm {\Theta}^f_{b}}$ in the TTD-RIS architecture is more complex, which increases the network scale of the DL-BF module in the E2E model. Due to the simplified network components, the convergence speed of the SA-RIS architecture is faster than the TTD-RIS architecture. However, with the increase of training epochs, the superior convergence performance can be obtained for the E2E model with TTD-RIS architecture.}

\begin{figure}[t]
	\centerline{\includegraphics[width=3.0in]{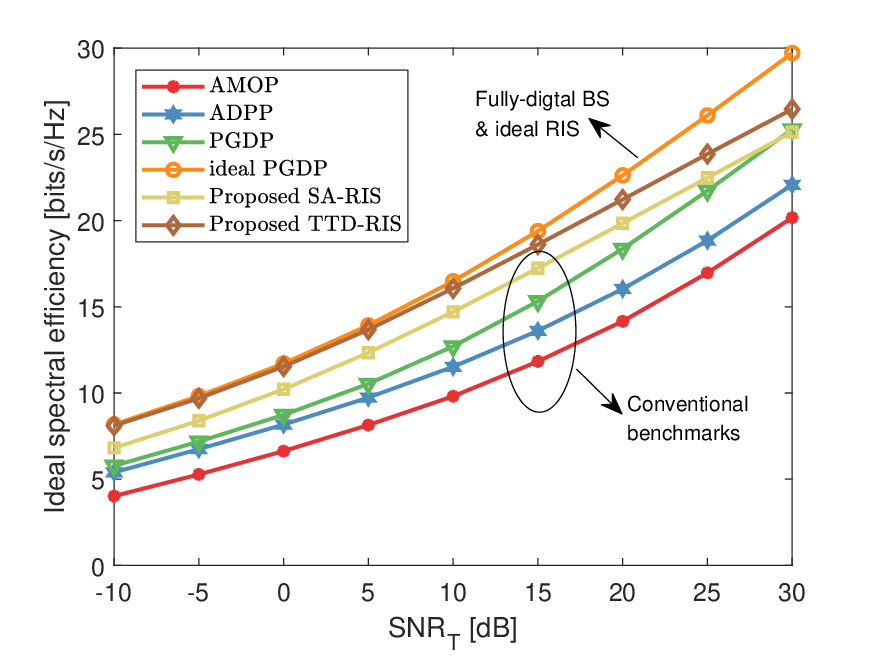}}
	\caption{Spectral efficiency versus downlink $\text{SNR}_\text{T}$ for different beamforming schemes, in which the perfect CSI is assumed to be known.}
	\label{R1}
\end{figure}

In Fig.~\ref{R1}, we present the spectral efficiency of the proposed E2E models and the ideal beamforming benchmarks with the perfect CSI. Note that for the proposed E2E models, i.e., the proposed TTD-RIS and SA-RIS in Fig.~\ref{R1}, the UL-CT module is exploited to learn the implicit CSI instead of the prior assumption of the perfect CSI. For the existing beamforming benchmarks, the near-field double beam split will degrade the efficient beamforming gain. Specifically, the AMOP method adopt the conventional hybrid beamformer architecture, which cannot design the frequency-dependent analog beamformer at the BS and the phase shifting at the RIS. In ADPP method, the BS is equipped with $M_\text{RF}K$ TTD units to construct the frequency-dependent hybrid-beamforming at the BS, while the PGDP method adopt the fully-digital precoding architecture at BS to avoid the wideband beam split at the BS. However, the wideband beam split at the RIS cannot be addressed pertinently for the ADPP and PGDP methods. Under the perfect CSI assumption, the ideal PGDP methods can obtain the performance upper bound, while the ideal RIS architecture cannot be implemented in practical communication systems. In the proposed E2E models, by exploiting frequency-dependent RIS architectures, i.e., the TTD-RIS and the SA-RIS, and developing deep learning-based beamforming networks, the beamforming gain of the proposed E2E models is superior to the conventional beamforming benchmarks. In the SA-RIS architecture, the effective array aperture will be shrunk due to the virtual subarray division strategy, and hence the frequency-dependent beamforming gain will be reduced compared to the TTD-RIS architecture. 

\begin{figure}[t]
	\centerline{\includegraphics[width=3.0in]{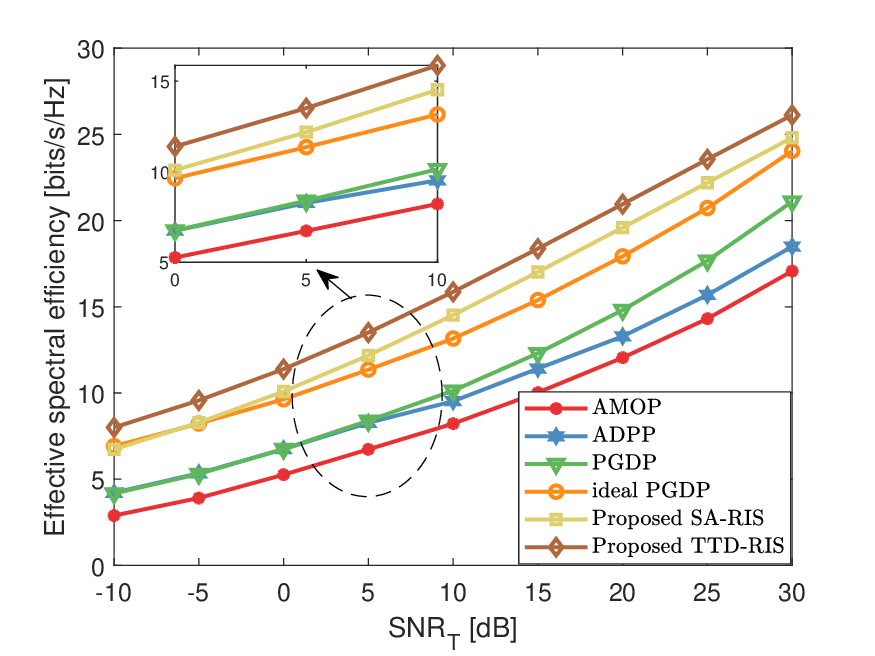}}
	\caption{Spectral efficiency versus downlink $\text{SNR}_\text{T}$ for different beamforming schemes, in which the uplink SNR is set to $\text{SNR}_\text{R}=10$ dB.}
	\label{R_BF}
\end{figure}

In Fig.~\ref{R_BF}, we further compare the effective spectral efficiency of the proposed E2E models with the practical beamforming benchmarks with the estimated CSI. Since the channel estimation error and the large pilot overhead, the effective spectral efficiency will be significantly decreased for the conventional beamforming schemes. {In the proposed E2E models with less pilot overhead, the available effective spectral efficiencies of both TTD-RIS and SA-RIS architectures are superior to the existing beamforming benchmarks.} 

\begin{figure}[t]
	\centerline{\includegraphics[width=3.0in]{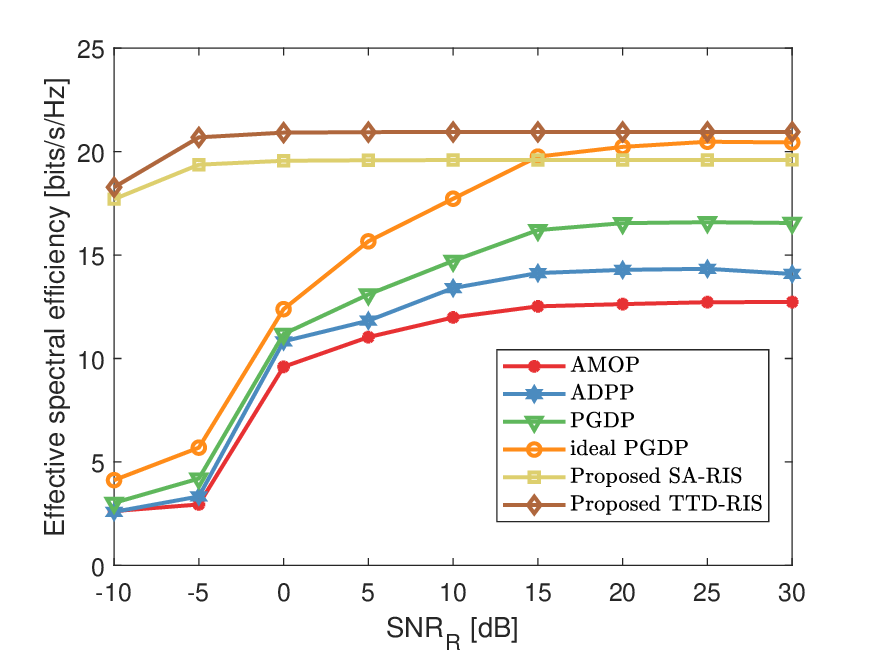}}
	\caption{Spectral efficiency versus uplink $\text{SNR}_\text{R}$ for different beamforming schemes, in which the downlink SNR is set to $\text{SNR}_\text{T}=20$ dB.}
	\label{R_CE}
\end{figure}

In Fig.~\ref{R_CE}, we compare the spectral efficiency of the proposed E2E models with the existing beamforming benchmarks under different uplink $\text{SNR}_\text{R}$. {For the case of low uplink SNR, the conventional beamforming approaches struggle as the noise-afflicted estimated CSI does not support efficient beamforming design. However, in the proposed E2E beamforming models, the implicit CSI acquisition and data-driven beamforming modules are jointly designed by constructing an efficient deep neural network, which can avoid the explicit error propagation between the channel estimation and beamforming modules in conventional approaches. Due to the ability of the network to learn specific latent representations from vast amounts of communication data, the proposed E2E models are robust against various disturbances in the input data. Moreover, the powerful nonlinear mapping capability of deep learning significantly reduces noise interference, offering a substantial improvement over the conventional approaches, particularly in low SNR conditions.} With the increase of $\text{SNR}_\text{R}$, the beamforming gain of all algorithms can be improved due to more accurate CSI, while the proposed E2E models are superior to the conventional beamforming benchmarks.
\begin{figure}[t]
	\centerline{\includegraphics[width=3.0in]{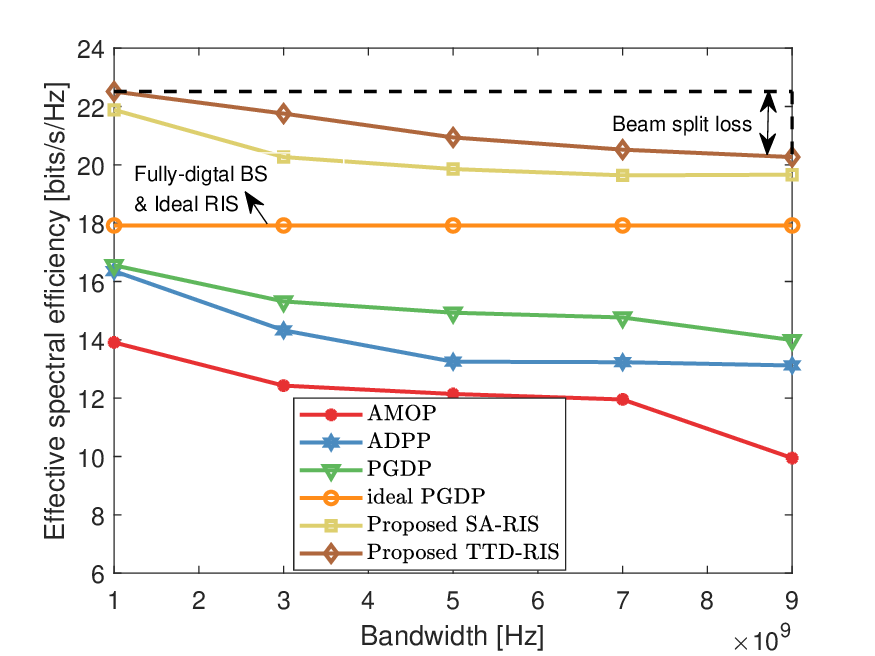}}
	\caption{Spectral efficiency versus bandwidth $W$ for different beamforming schemes.}
	\label{Band}
\end{figure}

{In Fig.~\ref{Band}, to characterize the beamforming performance loss caused by the near-field double beam split effect, we present the spectral efficiency of different beamforming schemes with increasing  bandwidth $W$. We observe a general degradation in the spectral efficiency of beamforming schemes that rely on the hybrid precoding architecture as $W$ increases, with the exception of the ideal PGDP algorithm implemented in a fully-digital architecture. In the ideal PGDP algorithm, the precoding matrices at the BS and the phase shifting at the RIS can be independently designed according to the specific subcarrier channel, achieving a beamforming gain that remains constant regardless of bandwidth $W$.} Compared to the existing conventional beamforming benchmarks, the proposed TTD-RIS and SA-RIS architectures have the superior generalization and resistance for the near-field double beam split effect under the large communication bandwidth.

\begin{figure}[t]
	\centerline{\includegraphics[width=3.0in]{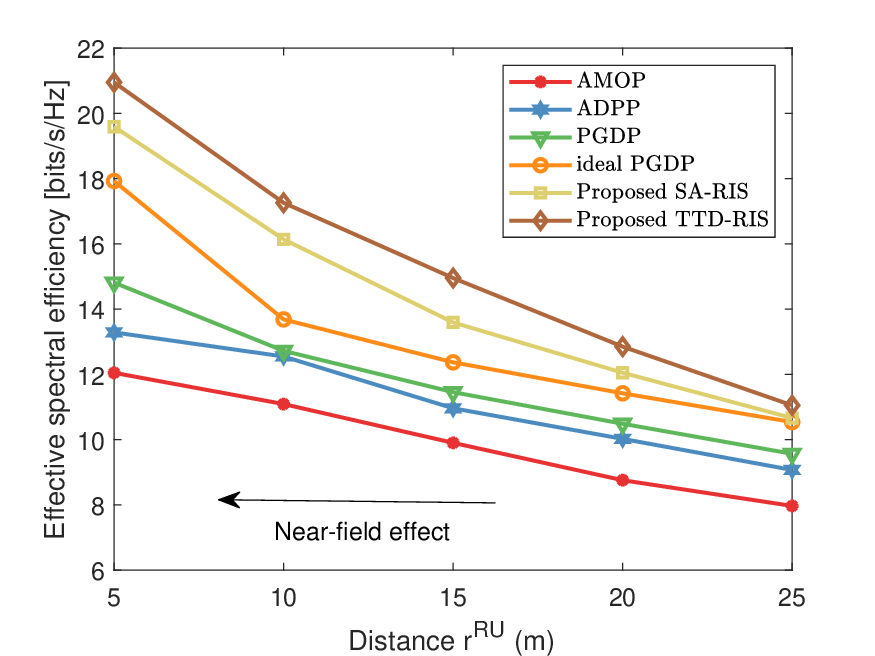}}
	\caption{{Spectral efficiency versus distance $r^\text{RU}$ for different beamforming schemes.}}
	\label{Distance}
\end{figure}

{In Fig.~\ref{Distance}, we present the performance comparison between different beamforming approaches as the near-field effect increases. Specifically, in the considered near-field RIS system, the closer the distance $r^\text{RU}$ between the RIS and the user, the more pronounced the near-field effect becomes. To accurately depict the near-field effect, we have normalized the large-scale fading components of each communication link to a reference value, highlighting the impact of distance-dependent array responses in near-field channel modeling. By leveraging the near-field effect of the virtual LOS channel introduced by the RIS, the spectral efficiency of all beamforming schemes improves as the distance $r^\text{RU}$ decreases. Compared to conventional beamforming approaches, the proposed beamforming models exhibit  superior performance gain as the near-field effect becomes increasingly significant, wherein the proposed TDD-RIS and SA-RIS architectures are specifically designed to effectively deal with the near-field double beam split effect.}


\subsection{{Comparison} for Different RIS Setups}

\begin{figure}[t]
	\centerline{\includegraphics[width=3.0in]{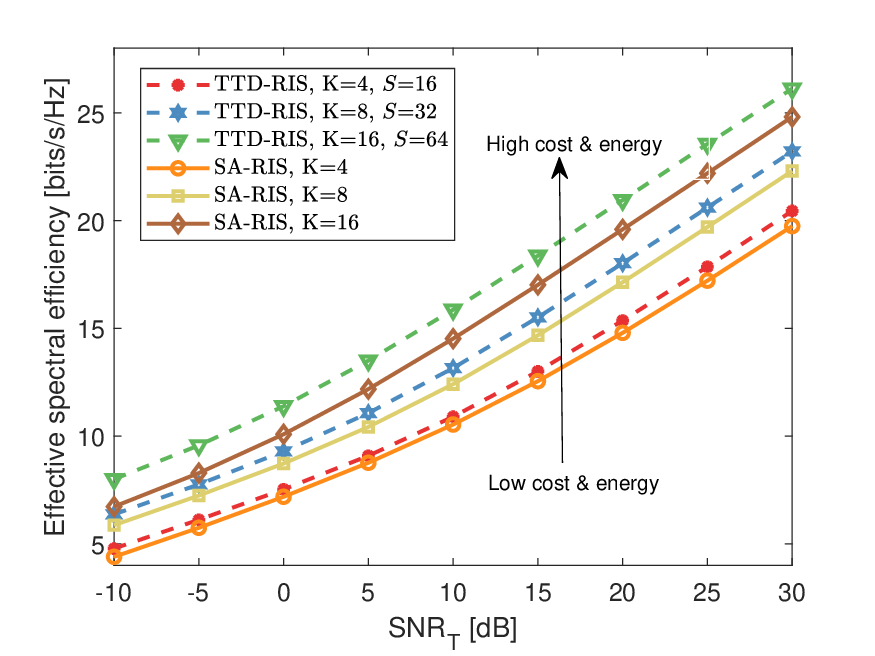}}
	\caption{Spectral efficiency versus downlink $\text{SNR}_\text{T}$ for different RIS setups.}
	\label{R_BF_P}
\end{figure}

\begin{figure}[t]
	\centerline{\includegraphics[width=3.0in]{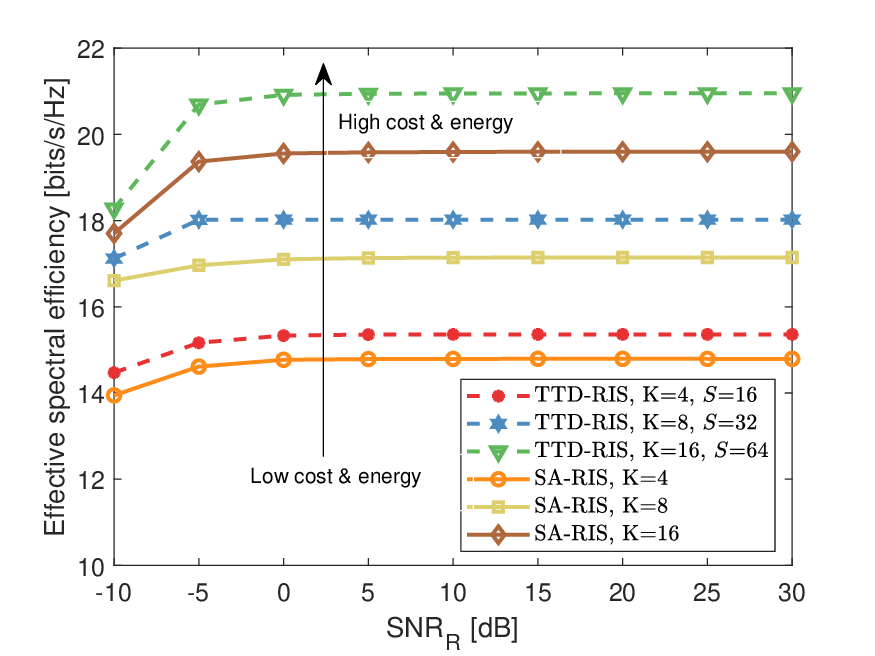}}
	\caption{Spectral efficiency versus uplink $\text{SNR}_\text{R}$ for different RIS setups.}
	\label{R_CE_P}
\end{figure}

In Fig.~\ref{R_BF_P} and Fig.~\ref{R_CE_P}, we present the spectral efficiency of the proposed E2E models under different RIS setups versus downlink and uplink SNR, respectively. With the increase of the number of TTD units $K$ at the BS, the beamforming performance of both TTD-RIS and SA-RIS architectures has been enhanced. For the TTD-RIS architecture, the spectral efficiency can be further upgraded by increasing the number of TTD units ${S}$ at the RIS, while the number of virtual subarrays at the SA-RIS architecture is fixed as the number of subcarriers $B$. Note that the introduction of TTD units and double-layer phase shifting circuits in the TTD-RIS architecture require the additional energy consumption and hardware cost. {Specifically, the typical power consumption of a TTD unit is about 100 mW [7], while a 3-bit PS only about 1.5 mW [8]. In the TTD-RIS architecture, with ${S} = 8 \times 8$ subarrays and ${P}^\text{S} = N \times 2 = 16 \times 32 \times 2$ PSs, the power consumption of the TTD-RIS can be calculated as ${S} \times 100 + {P}^\text{S} \times 1.5 = 7.936$ W. For the SA-RIS architecture without TTD units, the power consumption is $N \times 1.5 = 0.768$ W. Consequently, the SA-RIS architecture consumes significantly less power than the TTD-RIS architecture.}

\begin{figure}[t]
	\centerline{\includegraphics[width=3.0in]{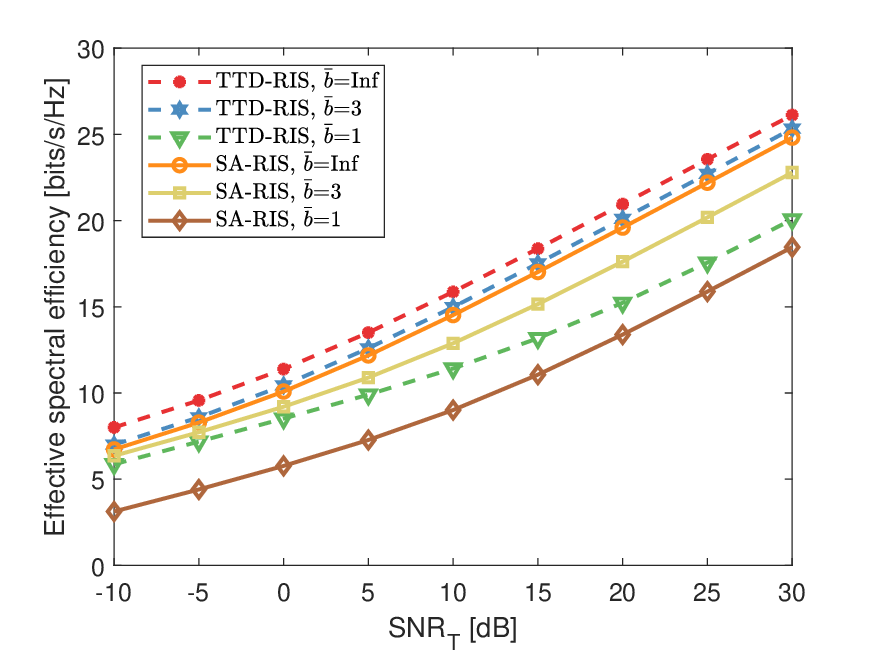}}
	\caption{{Spectral efficiency of the proposed E2E models under different phase shifting quantization bits $b$.}}
	\label{Bit1}
\end{figure}

{In Fig.~\ref{Bit1}, we provide the spectral efficiency of the proposed E2E models under different phase shift quantization bits $\bar{b}$. To realize the case of discrete phase shift, we incorporated quantization layers into both the UL-CT module and the DL-BF module, in which the quantized phase shift $\bar{\theta}$ with $\bar{b}$ bits for the original continuous phase shift $ {\theta} \in [0, 2\pi]$ can be expressed as $\bar{\theta} = \left\lfloor {\theta \times \frac{2^{\bar{b}}}{2\pi}} \right\rfloor \times \frac{2\pi}{2^{\bar{b}}}$. We employ a transfer learning strategy in \cite{Gao2022Data} to train the discrete E2E model with phase shift quantization modules. We observe that the beamforming performance of the proposed TTD-RIS and SA-RIS architectures, even with a modest $\bar{b}=3$ bits of phase shift quantization, closely approach the ideal performance of an infinite E2E model, which illustrates the robust generalization capability of the proposed E2E model in handling discrete phase shifts. In particular, in the case of low-resolution phase shifts, the TTD-RIS architecture with sub-connected TTD units consistently exhibits more stable performance compared to the SA-RIS architecture.}

\begin{figure}[t]
	\centerline{\includegraphics[width=3.0in]{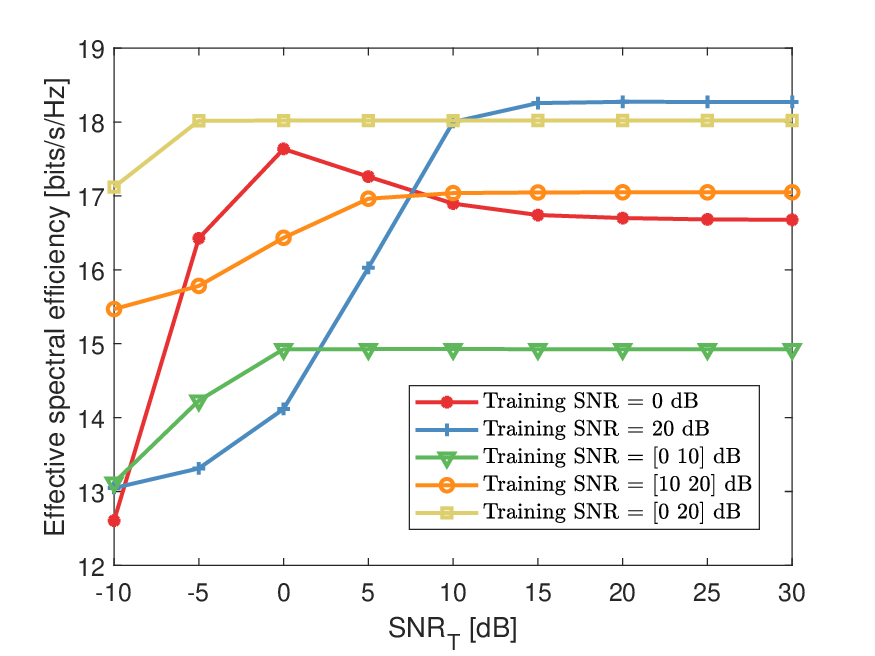}}
	\caption{{Spectral efficiency of the proposed TTD-RIS architecture versus downlink $\text{SNR}_\text{T}$ for different training SNR setups.}}
	\label{R_TSNR}
\end{figure}

In Fig.~\ref{R_TSNR}, we investigate the influence of the uplink SNR setting in the training stage for the proposed E2E models with the TTD-RIS architecture, in which the number of units at the BS and the RIS are set to $K = 8$ and ${S} = 32$, respectively. {We observe that the dynamic training SNR setting employed in this work, i.e., training $\text{SNR}_\text{R} \in [0,\ldots,20]$ dB, achieves more stable beamforming performance across the entire SNR range compared to other training SNR settings. This finding demonstrates that an appropriate training SNR setting can significantly enhance the testing performance of the deep learning-based beamforming model. In this work, the utilized training SNR setting takes into account the interference of communication noise and can also describe the effective distribution of data well.} In particular, for the trained model by using the fixed SNR setting, the model has satisfactory performance for the given SNR in the test stage, while the trained model lacks the robustness for the dynamic SNR range. In the dynamic training SNR setup, the training sample space is enriched by introducing different levels of noise components. {In essence, this training strategy {resembles} the data augmentation method in the traditional deep learning field, which can be extended to various deep learning-empowered communication scenarios, such as mmWave MIMO channel estimation \cite{10077727,9207745}.} 

\section{Conclusions}

In this paper, a deep learning enabled near-field wideband beamforming scheme in RIS-aided MIMO systems was proposed, aiming for alleviating the beamforming performance loss caused by the near-field double beam split effect. Firstly, two specific RIS architectures, i.e., TTD-RIS and SA-RIS, were exploited to achieve the frequency-dependent passive beamforming. Compared to the SA-RIS architecture, the TTD-RIS architecture can obtain superior beamforming performance, while requiring more energy consumption and hardware cost due to the introduction of TTD units. Furthermore, the E2E beamforming optimization framework was proposed to jointly design the high-dimensional channel estimation and the frequency-dependent wideband beamforming. Moreover, to accelerate the convergence of the proposed E2E model, the advanced deep learning architectures and the classical communication signal processing theory were integrated to develop an efficient beamforming network backbone. Numerical results showed the proposed E2E models without the explicit CSI had superior beamforming performance and robustness to the existing wideband beamforming benchmarks.

\ifCLASSOPTIONcaptionsoff
  \newpage
\fi

\bibliographystyle{IEEEtran}
\bibliography{IEEEabrv,refs_NFWB.bib}
\begin{IEEEbiography}[{\includegraphics[width=1in,height=1.25in,clip,keepaspectratio] {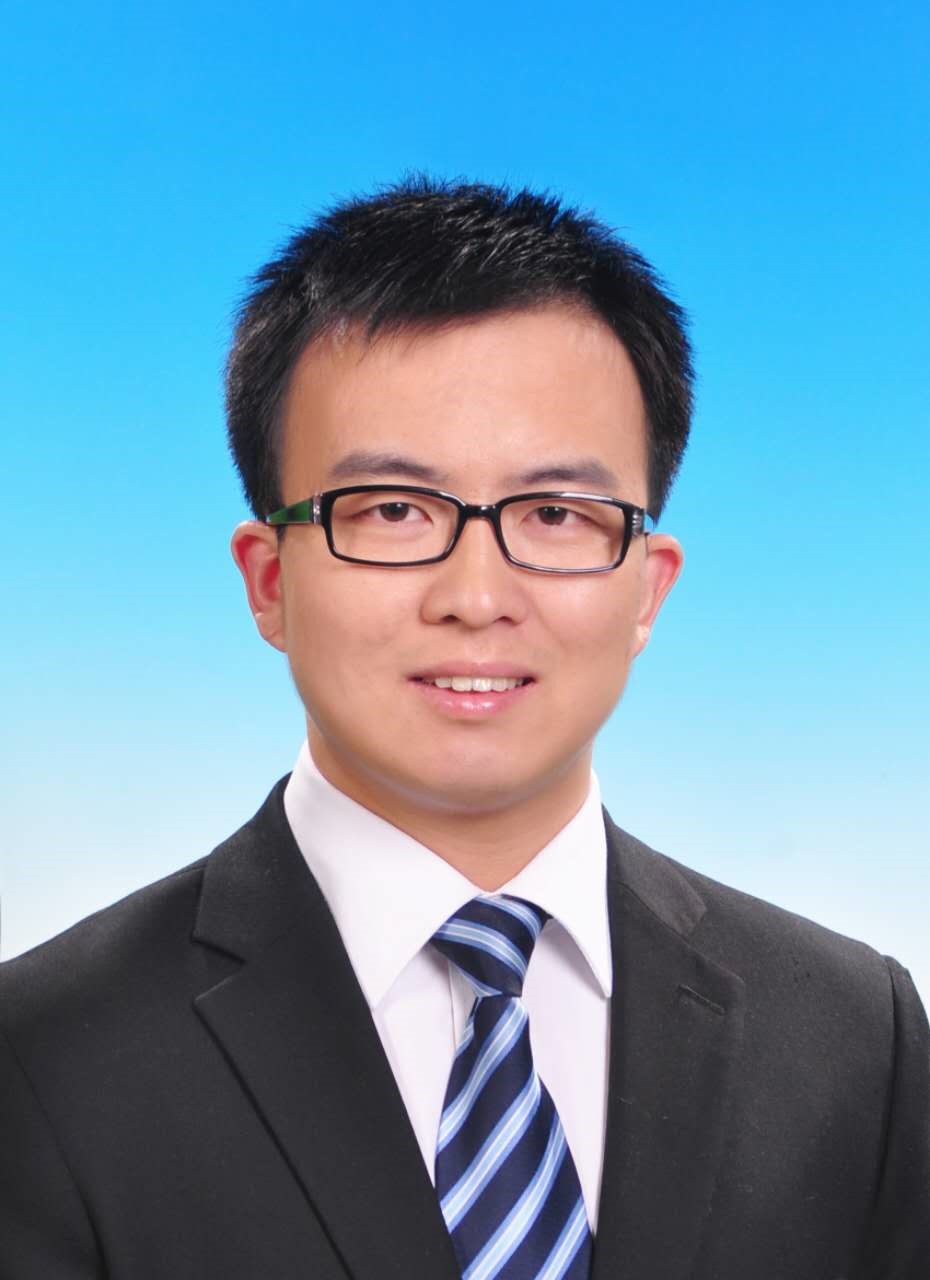}}]{Ji Wang} received the B.S. degree from the School of Electronic Information and Communications, Huazhong University of Science and Technology, China, in 2008, and the Ph.D. degree from the School of Information and Communications Engineering, Beijing University of Posts and Telecommunications, China, in 2013. He is currently an Associate Professor with the Department of Electronics and Information Engineering, Central China Normal University, China. Prior to that, he held postdoctoral positions with the School of Electronic Information and Communications, Huazhong University of Science and Technology, and the Department of Electrical Engineering, Columbia University, USA. His research interests include 5G/6G networks and machine learning. \end{IEEEbiography}

\begin{IEEEbiography}[{\includegraphics[width=1in,height=1.25in,clip,keepaspectratio]{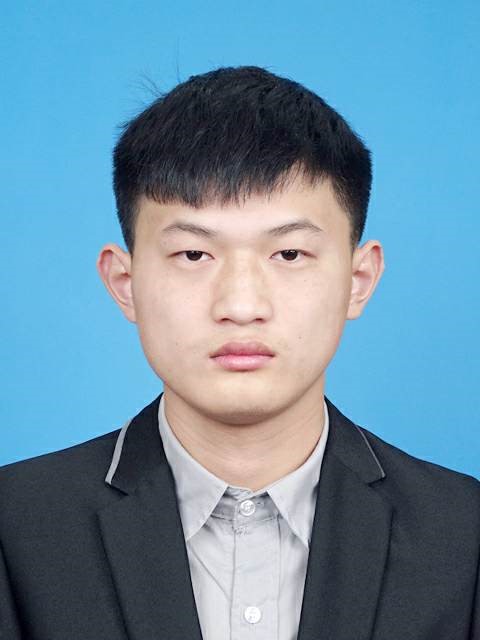}}]{Jian Xiao} received the B.Eng. degree and the M.Sc. degree (Distinction) from the Hunan Institute of Science and Technology, Yueyang, China, in 2019 and 2022, respectively. He is currently pursuing the Ph.D. degree with Central China Normal University. His research interests include electromagnetic information theory, reconfigurable intelligent surface, and machine learning.
\end{IEEEbiography}

\begin{IEEEbiography}[{\includegraphics[width=1in,height=1.25in,clip,keepaspectratio]{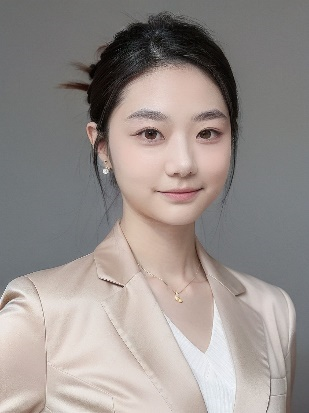}}]{Yixuan Zou} is a Lecturer (Assistant Professor) with the School of Electrical Engineering and Computer Science, Queen Mary University of London. She received the B.Sc. degree in Mathematics and the M.Sc. degree from Imperial College London, U.K., in 2017 and 2018, respectively. In 2022, she received her Ph.D. degree in Computer Science from Queen Mary University of London. Her research interests include artificial intelligence (AI) for wireless communications, non-orthogonal multiple access (NOMA), and IRSs/RISs aided communications for 6G networks. She is an academic Fellow at the Digital Environment Research Institute. She served as a Technical Program Committee member for IEEE VTC2023-Fall, VTC2024-Fall, MECOM 2024. She also served as the workshop and symposium session chairs for IEEE ICC 2023, GLOBECOM 2023, and ICC 2024. She serves as the Local Management Officer of NGMA-ETI 1st \& 2nd QMUL 6G Workshop.
\end{IEEEbiography}

\begin{IEEEbiography}[{\includegraphics[width=1in,height=1.25in,clip,keepaspectratio] {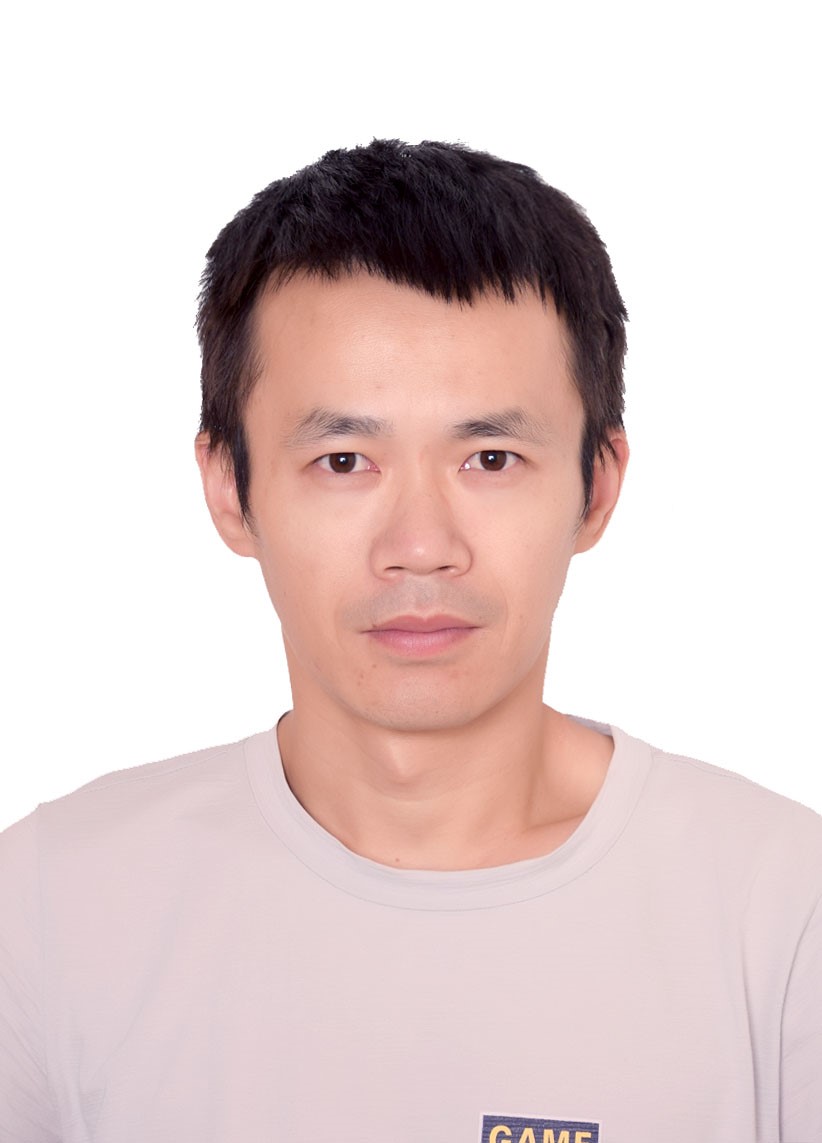}}]{Wenwu Xie} received the B.S., M.S., and Ph.D. degrees in communication engineering from Huazhong Normal University, Wuhan, China, in 2004, 2007, and 2017, respectively. He is currently an Associate Professor with the School of Information Science and Engineering, Hunan Institute of Science and Technology, Yueyang, China. His research interests include communication and control algorithms.
\end{IEEEbiography}

\begin{IEEEbiography}[{\includegraphics[width=1in,height=1.25in,clip,keepaspectratio] {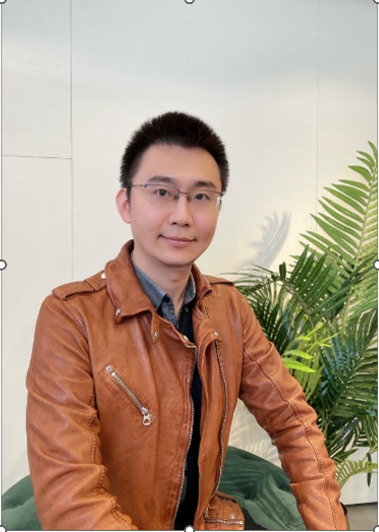}}]{Yuanwei Liu} (S'13-M'16-SM'19-F’24, \url{https://www.eee.hku.hk/~yuanwei/}) has been a (tenured) full Professor in Department of Electrical and Electronic Engineering (EEE) at The University of Hong Kong (HKU) since September, 2024. Prior to that, he was a Senior Lecturer (Associate Professor) (2021-2024) and a Lecturer (Assistant Professor) (2017- 2021) at Queen Mary University of London (QMUL), London, U.K, and a Postdoctoral Research Fellow (2016-2017) at King's College London (KCL), London, U.K. He received the Ph.D. degree from QMUL in 2016.  His research interests include non-orthogonal multiple access, reconfigurable intelligent surface, near field communications, integrated sensing and communications, and machine learning. 

Yuanwei Liu is a Fellow of the IEEE, a Fellow of AAIA, a Web of Science Highly Cited Researcher, an IEEE Communication Society Distinguished Lecturer, an IEEE Vehicular Technology Society Distinguished Lecturer, the rapporteur of ETSI Industry Specification Group on Reconfigurable Intelligent Surfaces on work item of “Multi-functional Reconfigurable Intelligent Surfaces (RIS): Modelling, Optimisation, and Operation”, and the UK representative for the URSI Commission C on “Radio communication Systems and Signal Processing”. He was listed as one of 35 Innovators Under 35 China in 2022 by MIT Technology Review. He received IEEE ComSoc Outstanding Young Researcher Award for EMEA in 2020. He received the 2020 IEEE Signal Processing and Computing for Communications (SPCC) Technical Committee Early Achievement Award, IEEE Communication Theory Technical Committee (CTTC) 2021 Early Achievement Award. He received IEEE ComSoc Outstanding Nominee for Best Young Professionals Award in 2021. He is the co-recipient of the 2024 IEEE Communications Society Heinrich Hertz Award, the Best Student Paper Award in IEEE VTC2022-Fall, the Best Paper Award in ISWCS 2022, the 2022 IEEE SPCC-TC Best Paper Award, the 2023 IEEE ICCT Best Paper Award, and the 2023 IEEE ISAP Best Emerging Technologies Paper Award. He serves as the Co-Editor-in-Chief of IEEE ComSoc TC Newsletter, an Area Editor of IEEE Communications Letters, an Editor of IEEE Communications Surveys \& Tutorials, IEEE Transactions on Wireless Communications, IEEE Transactions on Vehicular Technology, IEEE Transactions on Network Science and Engineering, IEEE Transactions on Cognitive Communications and Networking, and IEEE Transactions on Communications (2018-2023). He serves as the (leading) Guest Editor for Proceedings of the IEEE on Next Generation Multiple Access, IEEE JSAC on Next Generation Multiple Access, IEEE JSTSP on Intelligent Signal Processing and Learning for Next Generation Multiple Access, and IEEE Network on Next Generation Multiple Access for 6G. He serves as the Publicity Co-Chair for IEEE VTC 2019-Fall, the Panel Co-Chair for IEEE WCNC 2024, Symposium Co-Chair for several flagship conferences such as IEEE GLOBECOM, ICC and VTC. He serves the academic Chair for the Next Generation Multiple Access Emerging Technology Initiative, vice chair of SPCC and Technical Committee on Cognitive Networks (TCCN).
\end{IEEEbiography}
\end{document}